\documentclass[sigconf]{acmart}
\usepackage{booktabs} 
\usepackage{caption}
\usepackage{adjustbox}
\usepackage{subcaption}
\usepackage{relsize}
\usepackage{float}
\usepackage{url}
\usepackage{epstopdf}
\usepackage{graphics,graphicx}
\usepackage{balance}
\usepackage{array}
\usepackage{pifont}
\usepackage{multirow}
\usepackage{amsmath}
\usepackage{enumitem}
\usepackage{algorithm, algpseudocode}
\usepackage{footmisc}
\usepackage[normalem]{ulem}
\usepackage{paralist, tabularx}

\newcommand{\cmark}{\textcolor{blue}{\ding{51}}}%
\newcommand{\xmark}{\textcolor{red}{\ding{55}}}

\renewcommand\footnotetextcopyrightpermission[1]{} 

\makeatletter
\def\@copyrightspace{\relax}
\makeatother

\begin{document}
	\clubpenalty = 10000
	\widowpenalty = 10000
	
	\setlength{\belowdisplayskip}{0pt} 
	\setlength{\belowdisplayshortskip}{0pt}
	\setlength{\abovedisplayskip}{0pt} 
	\setlength{\abovedisplayshortskip}{0pt}
	
	\title{Alexa, in you, I trust! Fairness and Interpretability Issues in E-commerce Search through Smart Speakers}

	\author{Abhisek Dash}
	\affiliation{
		\institution{Indian Institute of Technology Kharagpur, India}
		\country{}
	}
	
	\author{Abhijnan Chakraborty}
	\affiliation{
		\institution{Indian Institute of Technology Delhi, India}
		\country{}
	}

	\author{Saptarshi Ghosh}
	\affiliation{
		\institution{Indian Institute of Technology Kharagpur, India}
		\country{}
	}
	
	\author{Animesh Mukherjee}
	\affiliation{
		\institution{Indian Institute of Technology Kharagpur, India}
		\country{}
	}
	
	\author{Krishna P. Gummadi}
	\affiliation{
		\institution{Max Planck Institute for Software Systems, Germany}
		\country{}
	}
	
	\renewcommand{\shortauthors}{Abhisek Dash et al.}

	\begin{abstract}
		In  traditional (desktop) e-commerce search, a customer issues a specific query and the system returns a ranked list of products in order of relevance to the query. An increasingly popular alternative in e-commerce search is to issue a {\it voice-query} to a smart speaker (e.g., Amazon Echo) powered by a {\it voice assistant} (VA, e.g., Alexa). In this situation, the VA usually spells out the details of \textit{only one} product, an \textit{explanation} citing the reason for its selection, and a \textit{default action} of adding the product to the customer's cart. This reduced autonomy of the customer in the choice of a product during voice-search makes it necessary for a VA to be far more responsible and trustworthy in its explanation and default action. 
		
		In this paper, we ask whether the explanation presented for a product selection by the Alexa VA installed on an \textit{Amazon Echo} device is consistent with human understanding as well as with the observations on other traditional mediums (e.g., desktop e-commerce search). Through a user survey, we find that in 81\% cases the interpretation of `a top result' by the users is different from that of Alexa. While investigating for the fairness of the default action, we observe that over a set of as many as 1000 queries, in $\approx$68\% cases, there exist one or more products which are more relevant (as per Amazon's own desktop search results) than the product chosen by Alexa. Finally, we conducted a survey over 30 queries for which the Alexa-selected product was different from the top desktop search result, and observed that in $\approx$73\% cases, the participants preferred the top desktop search result as opposed to the product chosen by Alexa. Our results raise several concerns and necessitates more discussions around the related fairness and interpretability issues of VAs for e-commerce search.
		\footnote{\textcolor{red}{This work has been accepted at The Web Conference 2022 (WWW'22). Please cite the version appearing in the conference proceedings.}}
	\end{abstract}

	
	
	\maketitle
	\pagestyle{empty}
	\section{Introduction}
Smart speakers like Amazon Echo or Google Nest have penetrated into the daily lives of millions across the globe, and are being increasingly used for varied purposes from making a phone call to purchasing something online~\cite{chung2017alexa,ram2018conversational, Mindstream2021Ways}. 
These smart speakers are powered by intelligent voice assistants (VA) like Alexa or Google Assistant. 
With such surge in VA usage, the research community has started looking into 
their impact in terms of privacy~\cite{chung2017alexa} and trustworthiness~\cite{nasirian2017ai, poushneh2021humanizing, foehr2020alexa}. 
In this work, we focus on one of the most important information access mechanisms -- e-commerce search using VAs (through smart speakers) -- and the 
consequences thereof.
We focus on e-commerce search because it is one of the most popular online activities in today's Web~\cite{IANS2021Indian,PTI2021Lockdown, Mindstream2021Ways}.

\begin{figure}[t]
	\centering
	\begin{subfigure}{0.45\columnwidth}
		\centering
		\includegraphics[width=\textwidth, height=3.2cm]{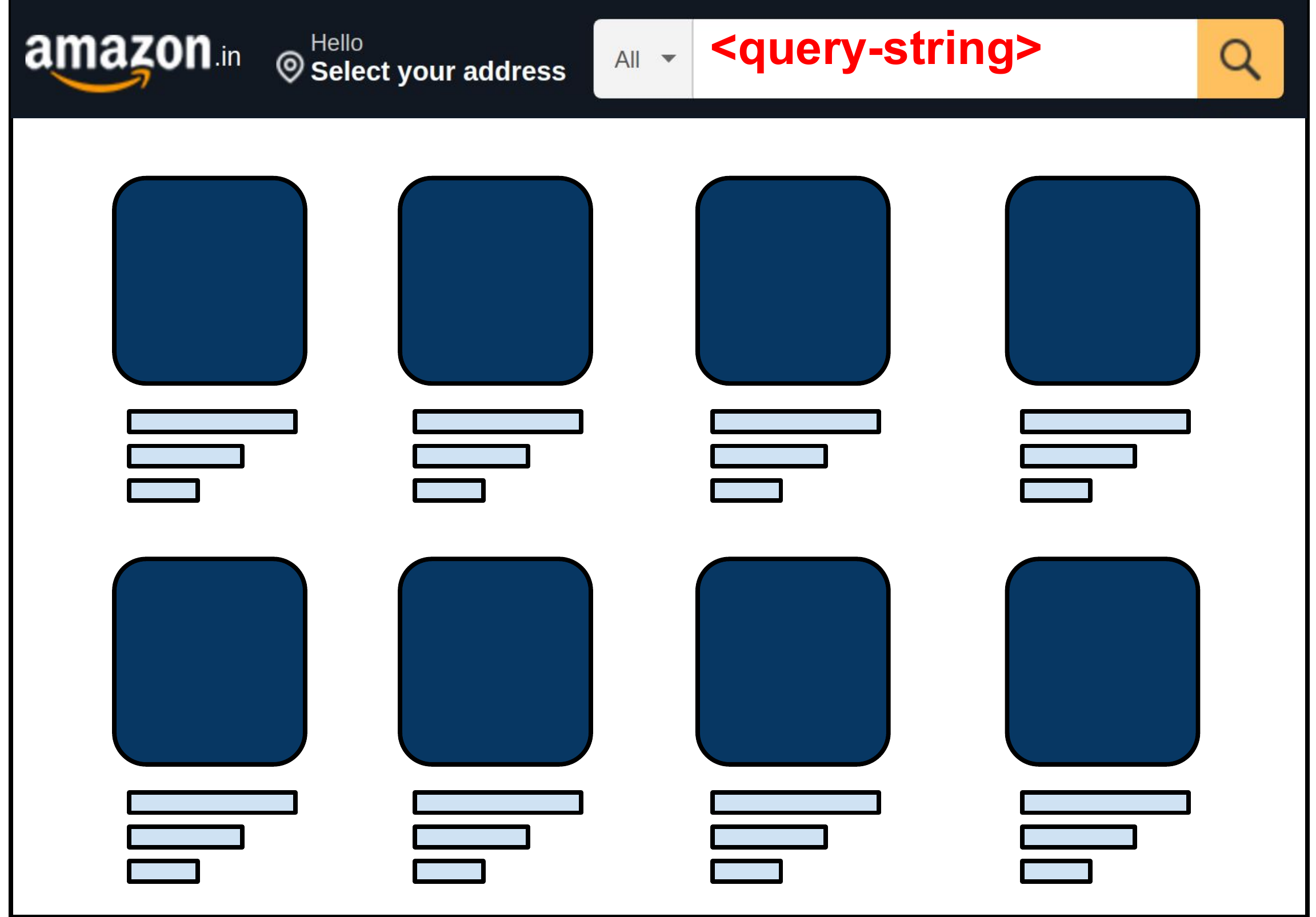}
		\caption{Desktop search on Amazon}
	\end{subfigure}
	\begin{subfigure}{0.53\columnwidth}
		\centering
		\includegraphics[width=\textwidth, height=3.2cm]{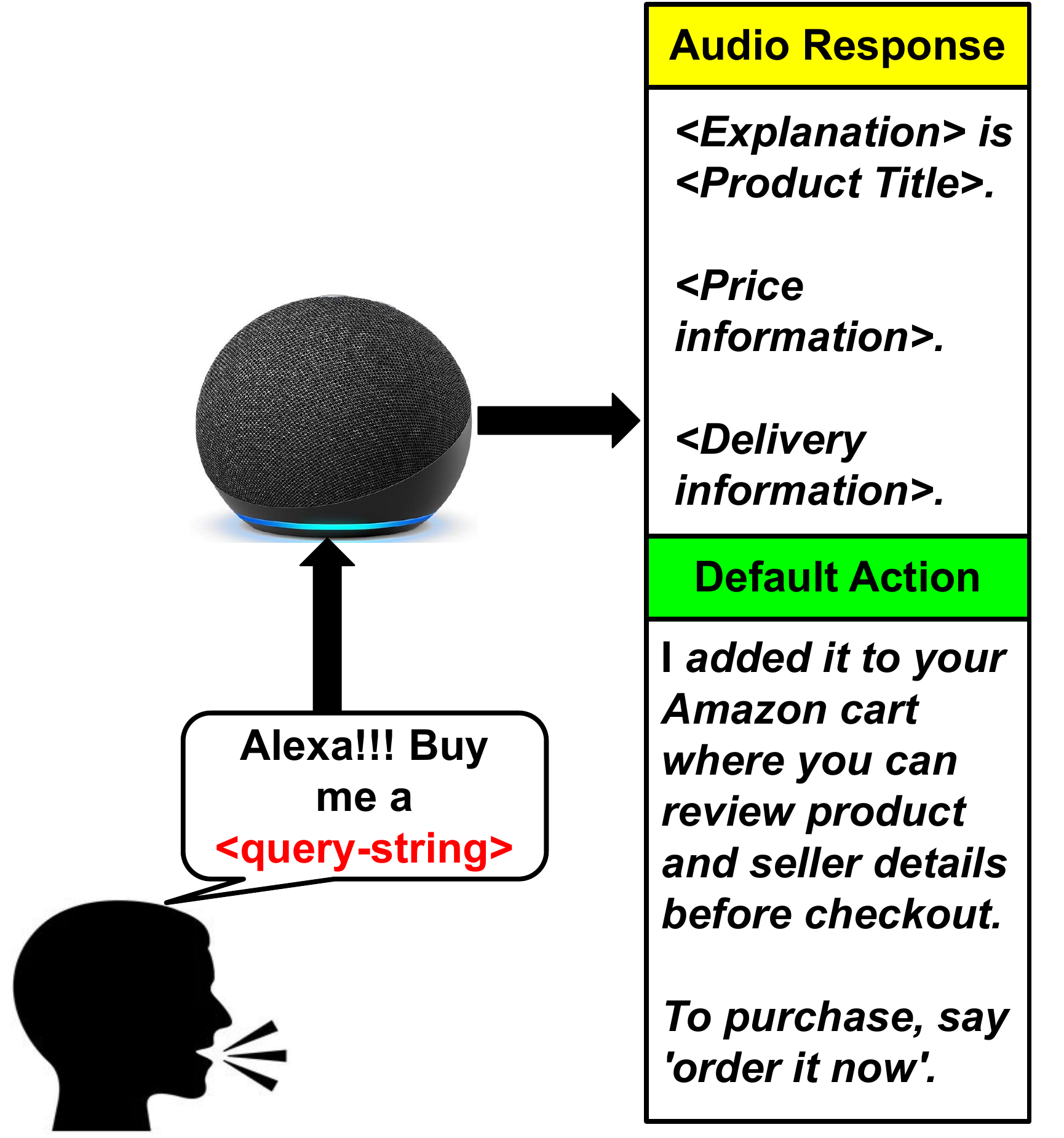}
		\caption{Search using Alexa}
	\end{subfigure}
	\vspace{-3 mm}
	\caption{ 
		(a)~In traditional desktop search, upon entering a query in the search bar a number of products are shown in a ranked order as per their relevance. (b)~In VAs through smart speakers, upon asking a purchase query (i)~VA spells out an audio response explaining a product information, (ii)~adds the aforementioned product to cart for further exploration.
	}
	\label{Fig: Search-Mediums}
	\vspace{-6 mm}
\end{figure}

\vspace{1 mm}
\noindent
\textbf{Traditional e-commerce search vs. search via VAs: }
In traditional e-commerce search, usually a user types a query in the search bar. A number of results, that the underlying search algorithm computes to be relevant, are shown in the decreasing order of relevance to the user.   
Figure~\ref{Fig: Search-Mediums}(a) shows a schematic diagram of 
the search paradigm in the Amazon e-commerce platform. The customer enters a query, and a ranked list of relevant products  
is shown along with some metadata (title, rating, price etc.). 
Given the abundance of choices, platforms are known to nudge users toward certain products by the 
way 
results are presented~\cite{baeza2018bias,schneider2018digital}. For example, results appearing at the top, 
toward the top-left corner of pages, or close to prominent images are known to accrue more clicks~\cite{baeza2018bias}.

In contrast, search through VAs\footnote{Note that throughout the rest of the paper, by search through VA we imply searching on a smart speaker which is powered by the VA as shown in Figure~\ref{Fig: Search-Mediums}(b).} do not offer such abundance of choices to the user. 
When a query is posed to a VA (e.g., Alexa), it typically responds with the details of {\it only one} product. 
Figure~\ref{Fig: Search-Mediums}(b) shows a prototype 
response (and Section~\ref{Sec: Data} gives several examples). 
The VA spells out the relevant product details (e.g., title, price and delivery information) and adds the product to the customer's cart for further review or purchase. In addition, it also 
asks the customer whether to make the purchase immediately. Often the response also contains a brief explanation of why 
the VA has chosen the corresponding product. 
For example, some prevalent explanations provided by Alexa VA is a product being the \textit{``Amazon's Choice''} or \textit{``a top result''}. 
Overall, we divide the response from VAs into two primary parts: 
(1)~an \textbf{audio response describing a chosen product with a brief explanation}, and 
(2)~a \textbf{status quo or default action} which is to add the chosen product to the customer's cart. 

Since a VA selects only a product for a given voice-query, customers cede complete autonomy to the decision making power of the VAs in such contexts. We posit that such restricted autonomy for the customers warrants the VAs to be more responsible and trustworthy in their response.
This situation raises two important concerns pertaining to the different parts of the response: (1)~How interpretable are the responses given by VAs to customers using the smart speakers?, and (2)~How fair are the default actions taken by VAs to different stakeholders involved in the process? 


\vspace{1 mm}
\noindent
\textbf{Interpretation of the explanation given by a VA: }
Users subconsciously extend the provided explanations 
with their own interpretations~\cite{kiesel2021meant, thaler2008nudge}. 
Such explanations (or responses) 
and their framing matter 
as users tend to make decisions passively, especially when there is a sense of urgency, such as in online purchase~\cite{kahneman2011thinking, thaler2008nudge}. 
Despite its importance, the interpretation of explanations from VAs has not been studied in the past. 
For example, how does a customer interpret an explanation such as a product is ``a top result'' or ``Amazon's Choice''?
Note that one can conduct a product search on multiple mediums nowadays, e.g., on a smart speaker, on Amazon's desktop website, or on its mobile app.
If the customers' interpretation of a VA's explanation is vastly different from what is observed in more traditional mediums (e.g., desktop search), it may affect their trust on 
the VAs in a negative way. 
Therefore, we posit that the consistency across results on different mediums is paramount for making the VAs more trustworthy.
This brings us to our first research question (\textit{RQ}) on \textbf{interpretability of the explanations provided by VAs}: \textit{\textbf{RQ--1}: How do users interpret the explanations given in the audio response by VAs?} 
More specifically, we intend to understand the users' interpretation of those explanations keeping traditional information access systems as a baseline. 

\vspace{1mm}
\noindent
\textbf{Fairness in the status quo action: } Humans have a general tendency to take the path of less effort, thus maintaining the status quo~\cite{thaler2008nudge}. 
Moreover, defaults have extra nudging ability because users tend to feel that they come with an implicit (or explicit) endorsement from the system~\cite{thaler2008nudge}. 
Thus if an option is designated as the default choice, 
the corresponding product can command a large market share~\cite{thaler2008nudge}. 
The likelihood of choosing the default option 
is further reinforced with explanations of positive sentiments. 
Such significant opportunity to revenue also brings with it several fairness concerns for both producers and customers. 
For example, in the context of e-commerce, \textit{non-selection of the most relevant product will deny its producers sale and revenue  opportunities} as well as \textit{mislead the customer to a (possibly) less relevant product} leading to customer dissatisfaction. 
This brings us to our second RQ on the \textbf{fairness of the default action} of a VA: \textit{\textbf{RQ--2}: How fair is the status quo action (product selection) of the VA?} 

\vspace{1 mm}
\noindent
\textbf{This study:} In this paper, we attempt to understand the aforementioned aspects of the explanation and the default action of Alexa VA to e-commerce search queries. 
The selection of Alexa is influenced by its popularity~\cite{Kinsella2020Voice, Perez2020Nearly} coupled with the vastness 
of Amazon as an e-commerce marketplace. Note that in our study the responses and the explanations within are both generated by Amazon Alexa. 

To this end, we created a scraper which automatically sends out e-commerce queries to Alexa (through an Amazon Echo device), and collects the product details of the product which was added to cart by Alexa. 
Further, we collect a snapshot of the {\it desktop Amazon search} result for the same query (keeping the search context as similar as possible, e.g., at the same time instant, from the same user account, same geographic location, same delivery location, etc.) to further analyze the customers' interpretations. 
Our selection of desktop Amazon search as a baseline is influenced by the popularity of the medium for online shopping~\cite{Tinuiti2020Amazon}. 
We also conducted a survey among 100 participants 
to understand their interpretations of the different explanations provided by Alexa VA while adding the products to their cart. 
By keeping the survey responses and the Alexa transcripts as references, we make the following observations.

\vspace{1 mm}
\noindent
$\bullet$ \textbf{Interpretation of explanations: }The interpretation of the respondents and the observations from the desktop search results taken immediately after the query was passed through Alexa do {\it not} align with each other in majority of the cases. In particular, we find that in \textbf{81\%} cases the interpretation of the survey participants about `a top result' does not match with that of Alexa. 

\noindent
$\bullet$ \textbf{(Un)fairness in the status quo action: }We observe that \textit{in \textbf{68\%} cases, one or more products were available in the desktop search result} which were more relevant than the product added to cart by Alexa.

\noindent
$\bullet$ \textbf{Preference of the customers: }Upon conducting a user survey for 30 queries (where the product added to cart by Alexa and the top search results differ), \textit{respondents preferred to buy the product at the top of search result to the one added by Alexa on \textbf{73\%} occasions}.

To the best of our knowledge, this is the first evaluation of its kind to understand users' interpretation of the responses provided by VAs (albeit for purchase queries only). We will make the dataset available upon request at: \url{https://forms.gle/aEG2n84Ay82QkVD19}. We believe that the insights drawn from this work will motivate development of more trustworthy VAs in future.
\section{Data collection}
\label{Sec: Data}
Next, we discuss the data collection process and different explanations obtained from Alexa for adding products to one's cart. 

\subsection{Data collection pipeline}
\begin{figure}[t]
	\centering
	\begin{subfigure}{0.8\columnwidth}
		\centering
		\includegraphics[width= \textwidth, height=3.4cm]{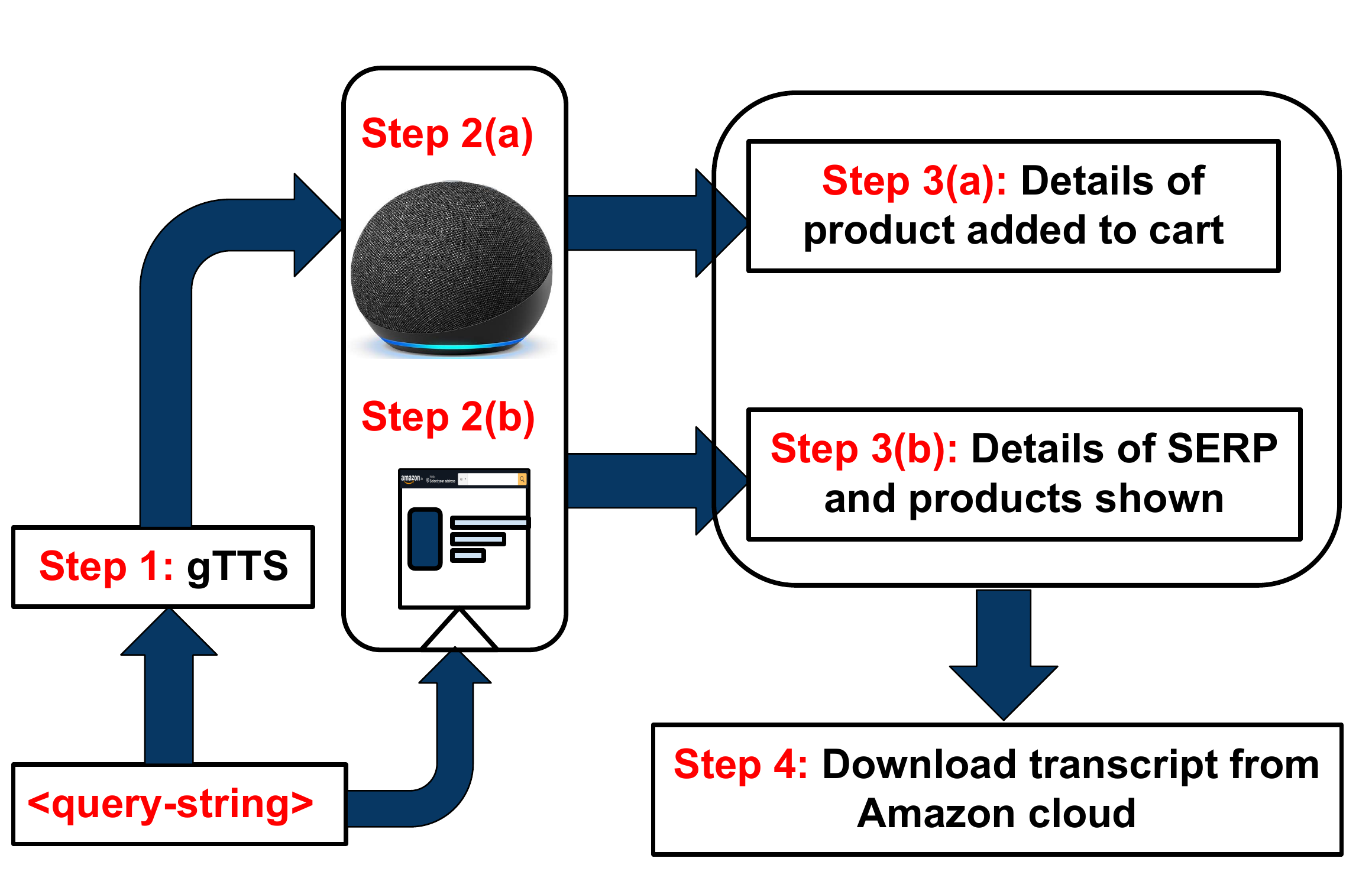}
	\end{subfigure}
	\vspace{-5 mm}
	\caption{The data collection pipeline. 
		A query string gets converted to an audio signal by Google Text-To-Speech. The audio signal and the textual query are provided to the Alexa VA and desktop search respectively. Product page details of the retrieved products and the transcripts of the Alexa conversation are then collected for further analyses.}
	\label{Fig: Alexa-Scraper}
	\vspace{-5 mm}
\end{figure}

Customers can interact with the Amazon search system through multiple mediums. In this work, to check the consistency of the Alexa selections and their explanations, we 
take Amazon desktop search as a baseline. Our choice of having desktop search as the baseline is influenced by the fact that it is the oldest and most popular medium of interaction of customers with Amazon~\cite{sorokina2016amazon}. Moreover, a 2020 survey suggest nearly 65\% Amazon shoppers use desktop website to shop on Amazon~\cite{Tinuiti2020Amazon}. 
Therefore, the data collection pipeline was set up in a way such that we ask a query to Alexa (through an Amazon Echo device) and perform a desktop search with the same query at the very same time instant on Amazon from the same user account. 
The data collection pipeline is shown in Figure~\ref{Fig: Alexa-Scraper} and the four major steps are explained below.

\noindent
\textbf{Step 1:} \textit{Text to speech conversion}: Given a query text, the first step is to generate a voice command for the Alexa voice assistant. We used Google Text-to-Speech (gTTS:  \url{https://gtts.readthedocs.io/en/latest/}) library to create automatic voice commands for the Alexa system. 

\noindent
\textbf{Step 2:} Search with the same query on Alexa device and Amazon desktop simultaneously.

\noindent
$\bullet$ Step 2(a): \textit{Search on Alexa}: We issue the voice-query "Alexa!!! buy me a $<$query-string$>$" using an \textit{Amazon Echo Dot (4th Gen)}. 

\noindent
$\bullet$ Step 2(b): \textit{Search on desktop}: As soon as the Alexa query was sent, we immediately searched for the same query on the Amazon website on a desktop. For desktop data collection, we performed browser automation using Gecko driver and Firefox web-browser. 

\noindent	
\textbf{Step 3:} Visit the product pages of the relevant products to collect meta-data and seller information. 

\noindent
$\bullet$ Step 3(a): The product added to cart by the Alexa VA.

\noindent
$\bullet$ Step 3(b): Products that appeared as desktop search results.

\begin{figure}[t]
	\centering
	\begin{subfigure}{0.48\columnwidth}
		\centering
		\includegraphics[width=\textwidth, height=3cm]{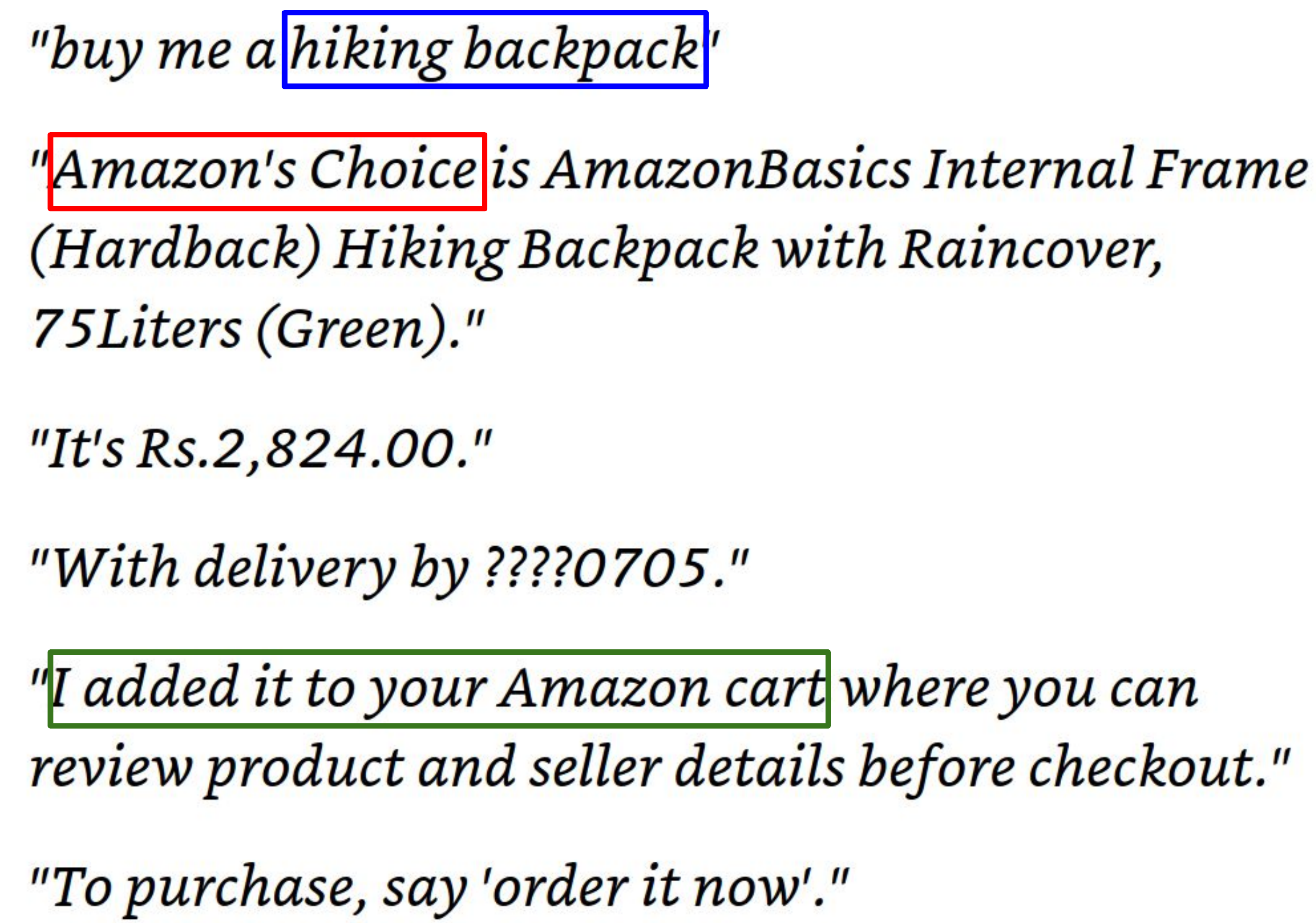}
		\caption{}
	\end{subfigure}
	\begin{subfigure}{0.48\columnwidth}
		\centering
		\includegraphics[width=\textwidth, height=3cm]{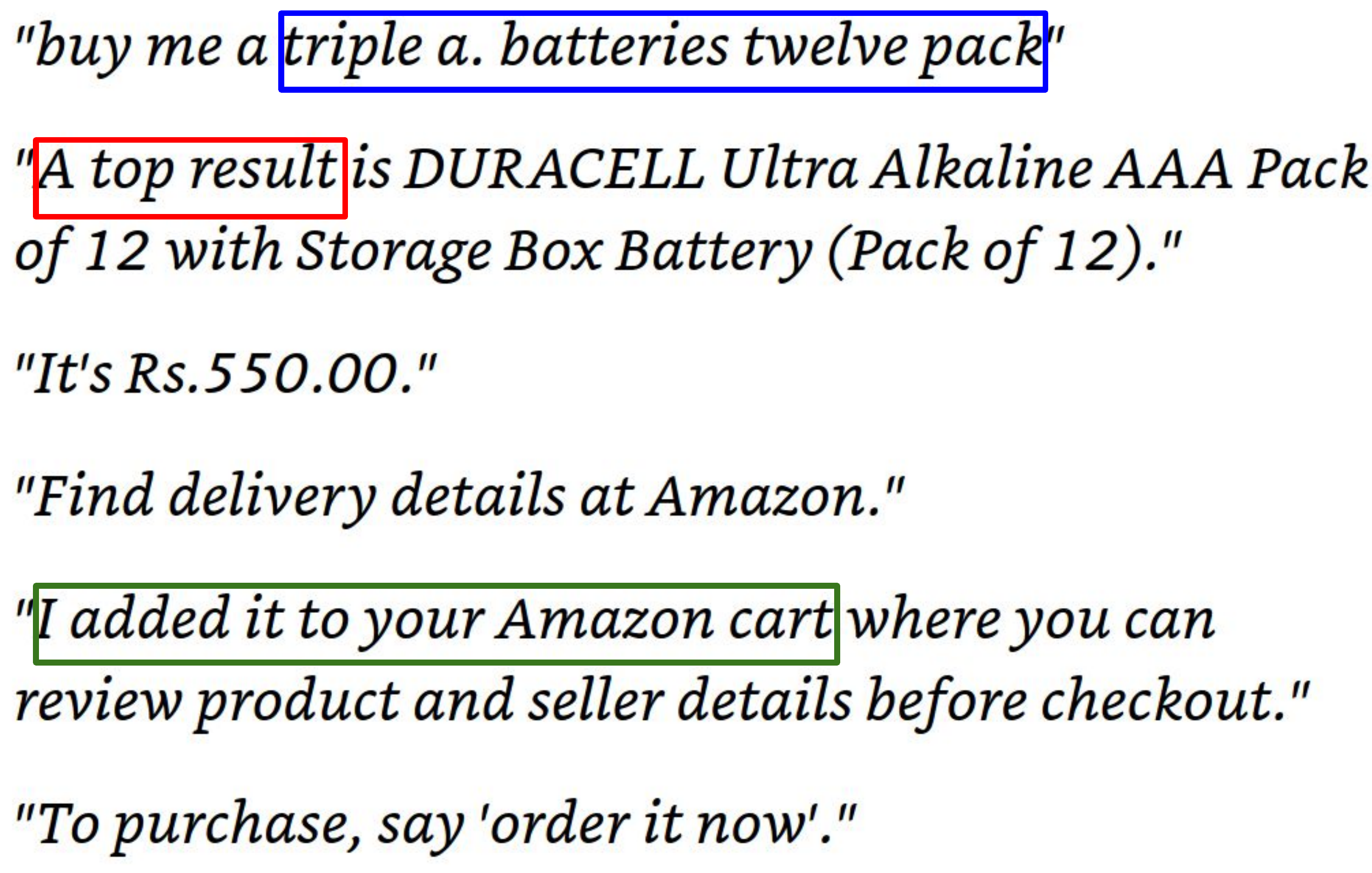}
		\caption{}
	\end{subfigure}
	\vspace{-4 mm}
	\caption{ Transcripts of responses given by Alexa for different queries (within blue rectangles). The explanation (within red rectangles) here is the corresponding product being (a)~the ``Amazon's Choice'', and (b)~`a top result'. The default action (within green rectangles) is that in both cases a product has been added to the customer's Amazon cart.}
	\label{Fig: Alexa-Transcripts}
	\vspace{-8 mm}
\end{figure}

\noindent	
\textbf{Step 4:} Download the Alexa transcript from Amazon cloud. This is done so that we can understand why Alexa added a product to the cart, i.e., to gather the explanation provided by Alexa. Some sample transcripts are shown in Figure~\ref{Fig: Alexa-Transcripts}.

\vspace{1mm}
\noindent
\textbf{Query selection: }
We first collected the top-100 most searched keywords on Amazon~\cite{Hardwick2021Top} and performed desktop search on Amazon using them. We also visited the product pages of products shown on the corresponding search engine result pages (SERPs) 
to obtain 
different popular queries mentioned on the product pages. These were included into the query list. Further, we also used Amazon's auto-complete suggestions to get additional queries. Our intention was to gather as many popular query strings as possible from the 100 keywords that we started with. 
Finally, we curated a list of 1000 keywords for which we collected the data using the above pipeline (Figure~\ref{Fig: Alexa-Scraper}). The query set comprises of queries from different popular product categories on Amazon, e.g., \textit{Electronics, Computer Accessories, Mobile Accessories, Home and Kitchen} 
etc. (see Table~\ref{Tab: QueryStat} for the top-10 categories in our query set).
Further, we also collected {\it 14 temporal snapshots of search results for each of the 100 primary keywords} to observe the temporal variations in results.  

\vspace{1mm}
\noindent
\textbf{Uniformity of the data collection process:} We collected all data using the aforementioned method from a single \textit{account having prime membership} in \textit{Amazon's Indian marketplace} (Amazon.in). 
We collected all the data from the same geographic location, using the same IP address and with the same delivery address to maintain uniformity. 
Further, to keep our analyses meaningful and comparisons fair, we performed the searches on both the devices (Echo and desktop) at the exact same time instant. 

We also used the temporal snapshots 
to check for stability of the top--k search results and the products selected to cart by Alexa. 
We observe that the selection of Alexa and the desktop search results are stable over time -- more than 6 products are retained in the top--10 desktop search results for 88\% of the queries in any consecutive pair of temporal snapshots, suggesting minimal stochasticity in the collected data. 
This last observation is particularly important since it presents the necessary evidence that all the results that we subsequently present in the paper are non-stochastic outcomes.
Readers can refer to Section~\ref{Sec: Sup-temporal} (Figure~\ref{Fig: Stability}) for more details.

\begin{table}[t]
	\noindent
	\small
	\centering
	\begin{tabular}{ |p{2.9 cm}|r||p{2.3 cm}|r|}
		\hline
		{\bf Category} & \# Query & {\bf Category} & \# Query\\
		\hline
		Electronics & 238 & 	Mobile accessories & 74\\
		\hline
		Computer accessories & 211 & Home improvement & 50 \\
		\hline
		Home \& Kitchen & 118 & Office products & 48\\
		\hline
		Health \& personal care & 112 & Video games & 25\\
		\hline
		Sports, fitness \& outdoors & 21 & Luggage \& bags & 79 \\
		\hline

	\end{tabular}	
	\caption{{\bf Break up of queries from top-10 categories present in our dataset as defined by Amazon. 
	}}
	\label{Tab: QueryStat}
	\vspace{-10 mm}
\end{table}

\subsection{Responses from Alexa voice assistant}

As shown in Figure~\ref{Fig: Search-Mediums}(b), responses provided by Alexa comprises two different parts: (a)~an audio response describing a selected product with an explanation, and (b)~a default action - adding the selected product to cart. 
For example, Figure~\ref{Fig: Alexa-Transcripts} shows the exact transcript of such responses. 
In Figure~\ref{Fig: Alexa-Transcripts}~(a) upon being asked to buy a `hiking backpack', Alexa selects a product to add to the customer's Amazon cart. First it responds with the product title, price and delivery details. This response is preceded with a brief explanation (e.g., `Amazon's Choice' in Figure~\ref{Fig: Alexa-Transcripts}~(a)).
Then, Alexa says \textit{`I added it to your Amazon cart'}. 
After this, the selected product can be reviewed and checked out from the Amazon cart of the customer. 
Finally, in case the customer wants to go ahead with the purchase, then they need to say \textit{`order it now'}. 
Notice that, through this default action, Alexa makes it significantly easier for a customer to review or purchase the selected product.

\vspace{1 mm}
\noindent
\textbf{Explanations for the product selection: }Often these audio responses also mention the reason for the selection of the product in the form of a small explanation in the beginning of the response. For example, in Figure~\ref{Fig: Alexa-Transcripts}, the corresponding products were added to the cart for being an \textbf{``Amazon's choice''} or \textbf{``a top result''} respectively. 
During our data collection, we found the following types of explanations in the response of Alexa to our queries. 

\vspace{1mm}
\noindent
\textbf{Based on Amazon's Choice:} Amazon's choice is the most prevalent explanation given by Alexa while adding a product to the cart for a query. For 662 out of the 1000 queries, Alexa added a product to the cart which is explained as being \textit{``Amazon's Choice''}. 
Transcript of one such response is shown in Figure~\ref{Fig: Alexa-Transcripts}(a). Note that according to Amazon, \textit{``Amazon's Choice highlights highly rated, well-priced products available to ship immediately''}~\cite{Amazon2021Choice}.

\vspace{1mm}
\noindent
\textbf{Based on top result:} The second most prevalent explanation  provided by Alexa for adding a product is being `a top result'. An example of such a response is shown in Figure~\ref{Fig: Alexa-Transcripts}(b). For 251 out of the 1000 queries, Alexa added a product to the cart which is explained as being \textit{``a top result''}.

\vspace{1mm}
\noindent
\textbf{Other explanations}: We find the above two explanations to be the most prevalent and they cover more than 91\% of all the product searches we performed. 
Apart from these, a few other explanations provided by Alexa include (i)~`the best selling option', (ii)~`based on order history' (when the user has searched with the same query in the past), (c)~`closest I can find', etc. 
We do not discuss these explanations in detail for brevity. However, some aggregate level results for the same have been reported in the supplementary material.

In the rest of this paper, we shall try to understand the interpretations of the two most prevalent explanations. 
Throughout, we will compare the Alexa result for a particular query with the desktop search result for the same query, fired almost immediately as the query fired on the Alexa VA (as described in Step 2). 
For brevity, we describe the observations on the snapshot of 1000 queries throughout this paper. The results from the additional temporal snapshots are added in the supplementary material.
	\section{Interpretation of explanations}\label{Sec: Interpretation}
\begin{figure}[t]
	\centering
	\begin{subfigure}{0.48\columnwidth}
		\centering
		\includegraphics[width= \textwidth, height=2.5cm]{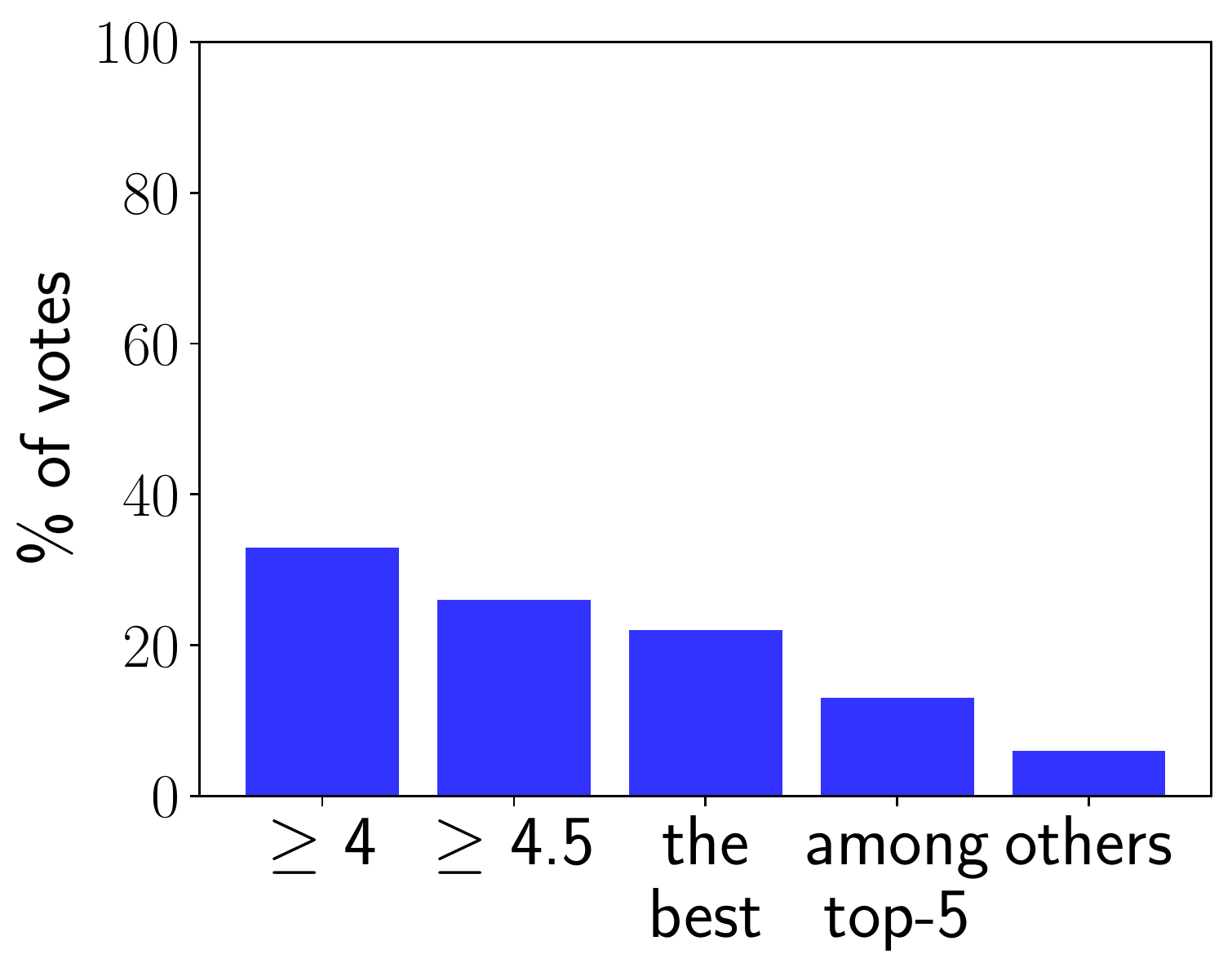}
		\vspace{-6 mm}
		\caption{AC -- Rating}
	\end{subfigure}
	\begin{subfigure}{0.48\columnwidth}
		\centering
		\includegraphics[width= \textwidth, height=2.5cm]{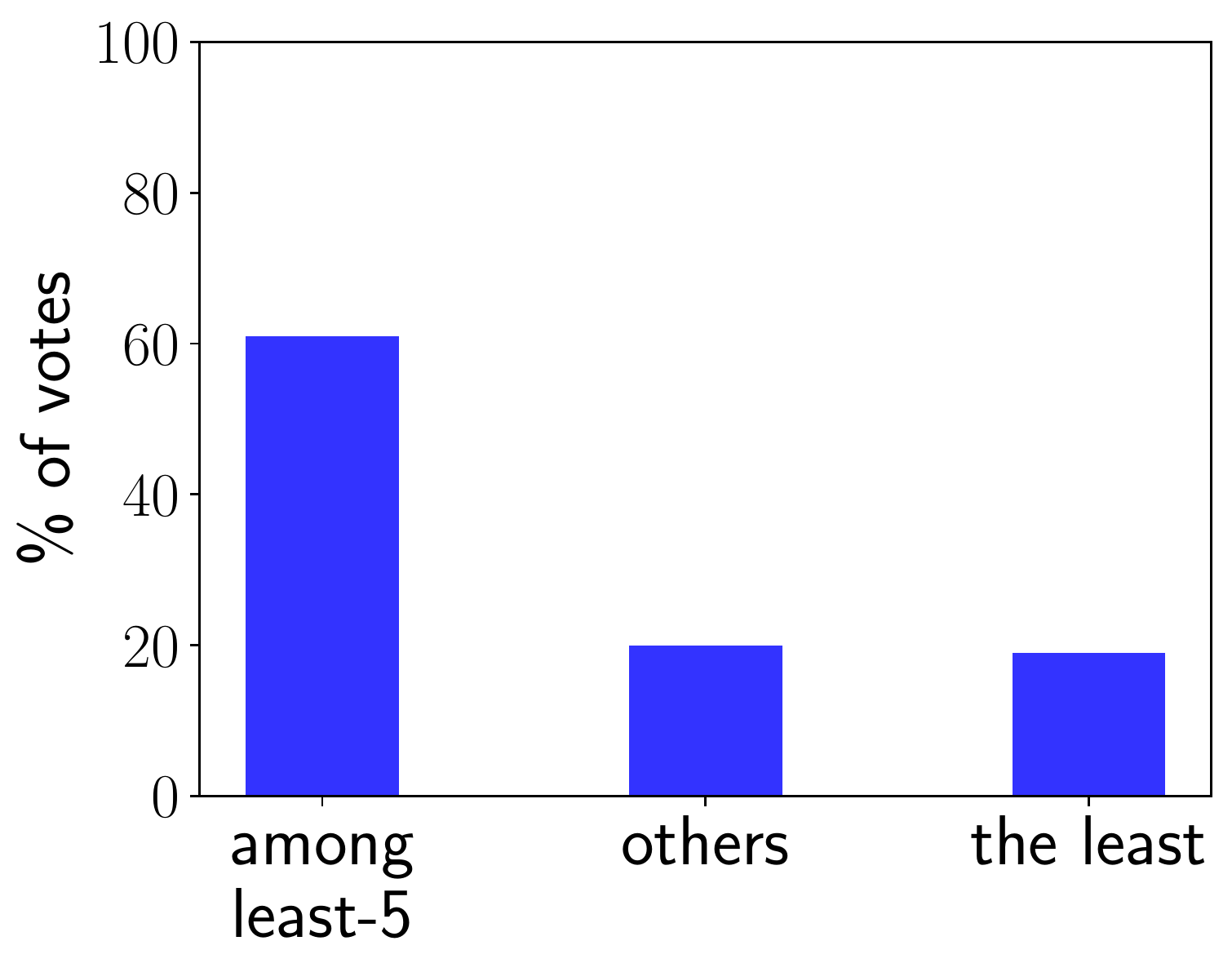}
		\vspace{-6 mm}
		\caption{AC -- Price}
	\end{subfigure}
	\begin{subfigure}{0.48\columnwidth}
		\centering
		\includegraphics[width= \textwidth, height=2.5cm]{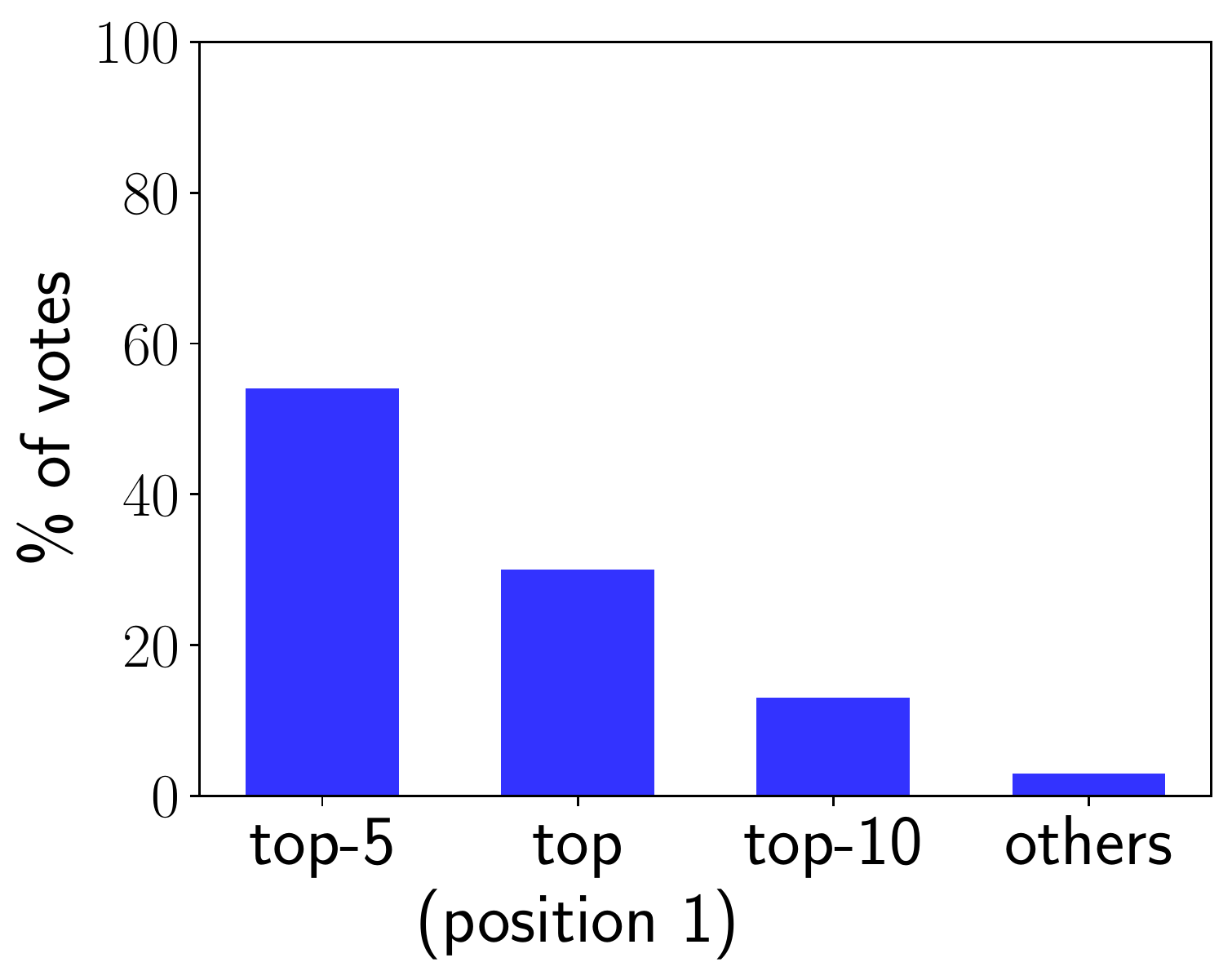}
		\vspace{-6 mm}
		\caption{AC -- Position}
	\end{subfigure}
	\begin{subfigure}{0.48\columnwidth}
		\centering
		\includegraphics[width= \textwidth, height=2.5cm]{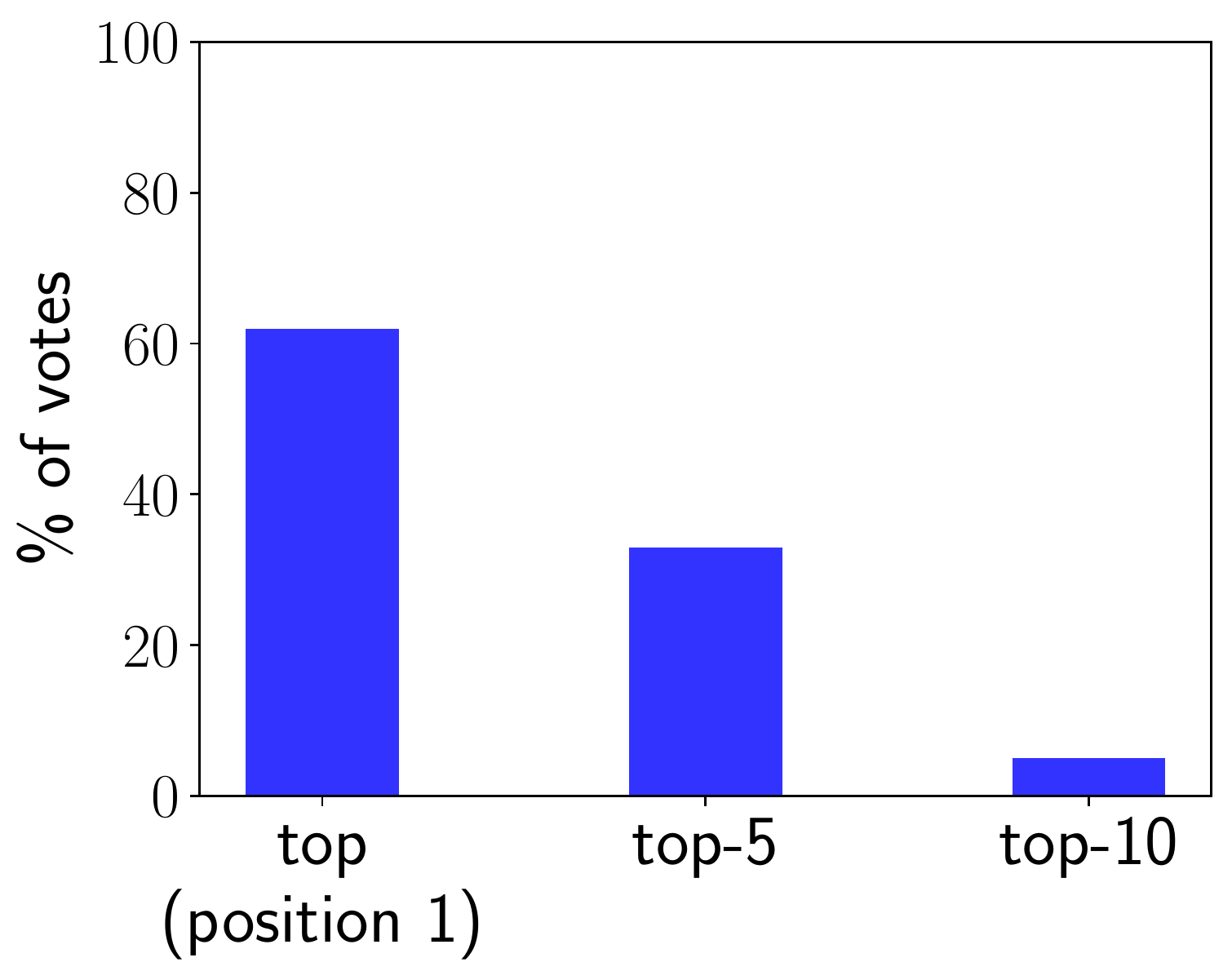}
		\vspace{-6 mm}
		\caption{Top -- Position}	
	\end{subfigure}
	\begin{subfigure}{0.48\columnwidth}
		\centering
		\includegraphics[width= \textwidth, height=2.5cm]{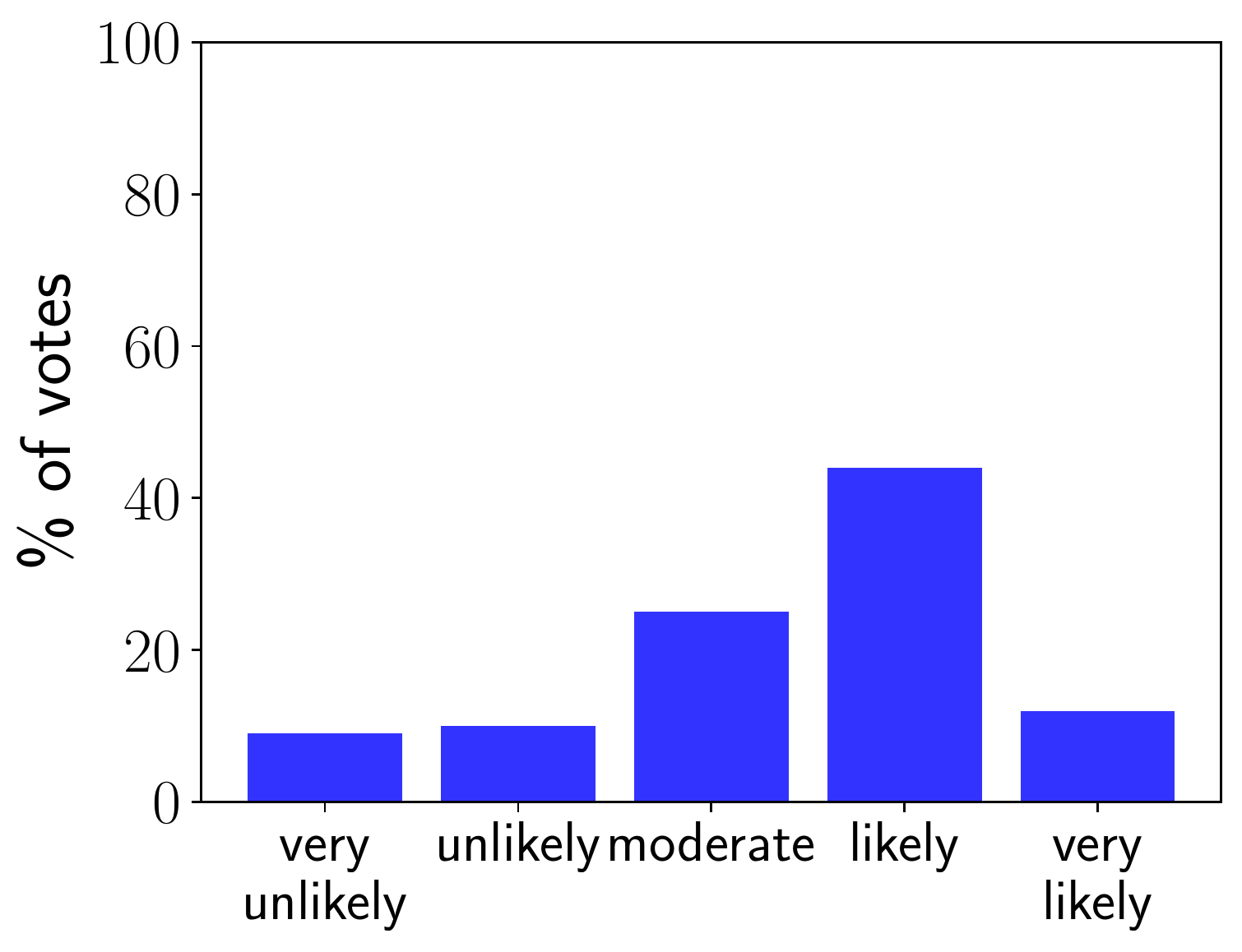}
		\vspace{-6 mm}
		\caption{AC -- Purchase likelihood}
	\end{subfigure}
	\begin{subfigure}{0.48\columnwidth}
		\centering
		\includegraphics[width= \textwidth, height=2.5cm]{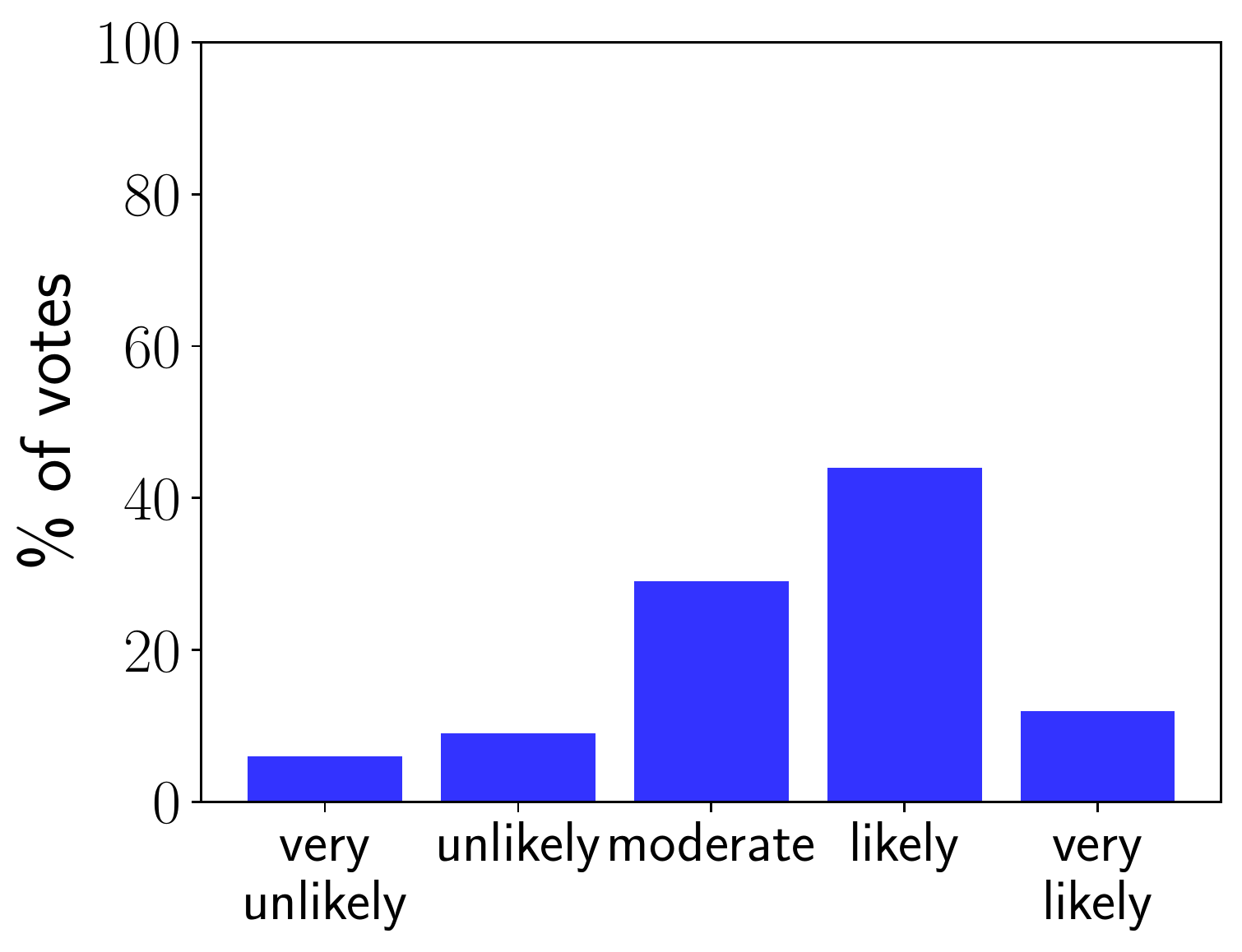}
		\vspace{-6 mm}
		\caption{Top -- Purchase likelihood}
	\end{subfigure}
	\vspace{-4 mm}
	\caption{Bar plots of responses 
		regarding the interpretation of two most prevalent explanations of Alexa, i.e., `Amazon's Choice' (a--c), and `a top result'' (d). Figures (e), (f): Responses of purchase likelihood of each explanation type.}
	\label{Fig: Survey-Interpretation-Responses}
	\vspace{-5 mm}
\end{figure}

To understand the interpretation of the explanations provided by Alexa system, we conducted a survey among 100 participants. A majority of our respondents are male (72\%) and in the age group of 20--30 years. All of them are very conversant with shopping on Amazon. In this section, 
we present the interpretation from the survey and then compare the same with the observations in the corresponding desktop search results.

\begin{figure*}[t]
	\centering
	\begin{subfigure}{0.6\columnwidth}
		\centering
		\includegraphics[width= \textwidth, height=3cm]{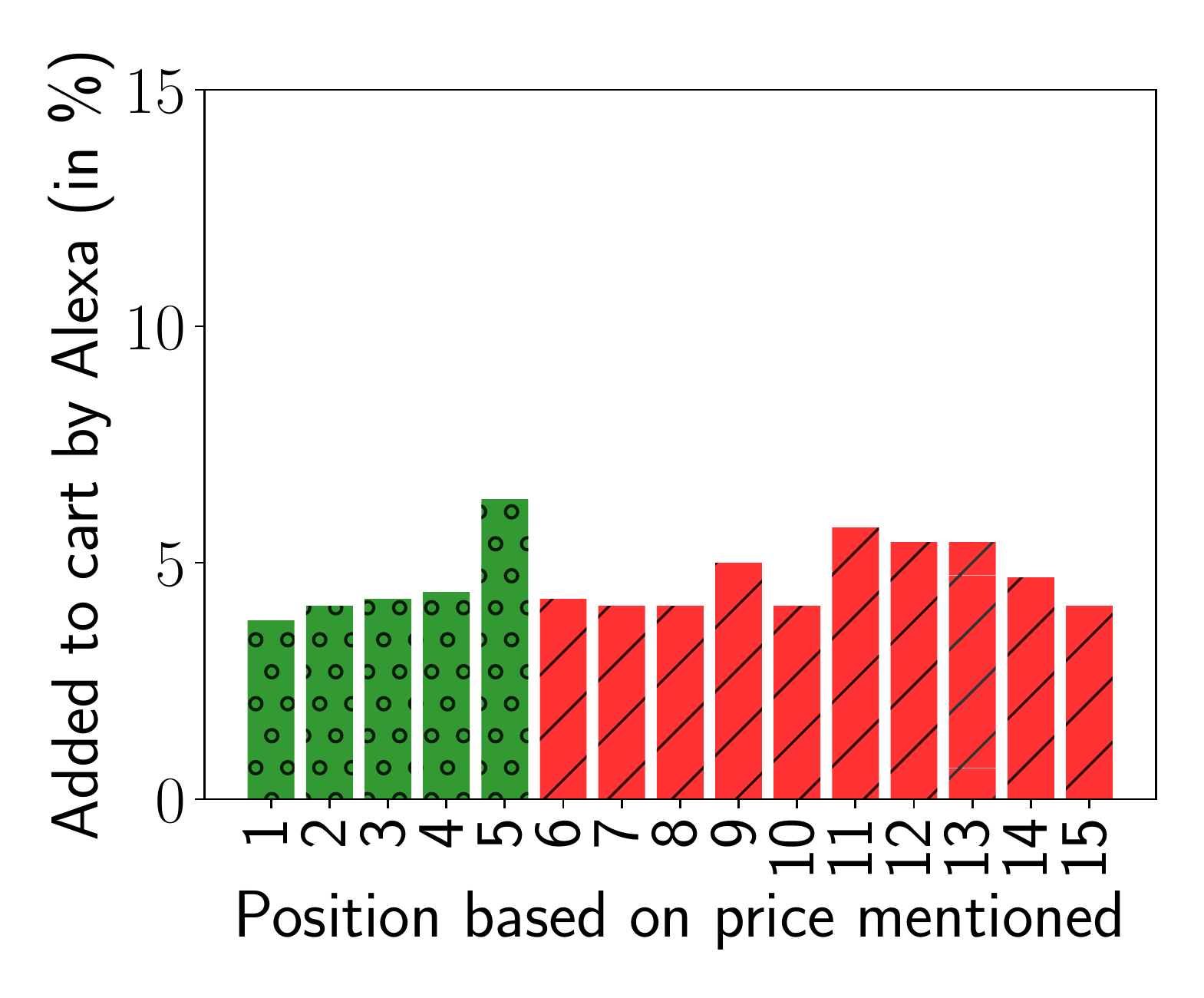}
		\vspace{-4 mm}
		\caption{Amazon's Choice -- Price}
	\end{subfigure}
	\begin{subfigure}{0.6\columnwidth}
		\centering
		\includegraphics[width= \textwidth, height=3cm]{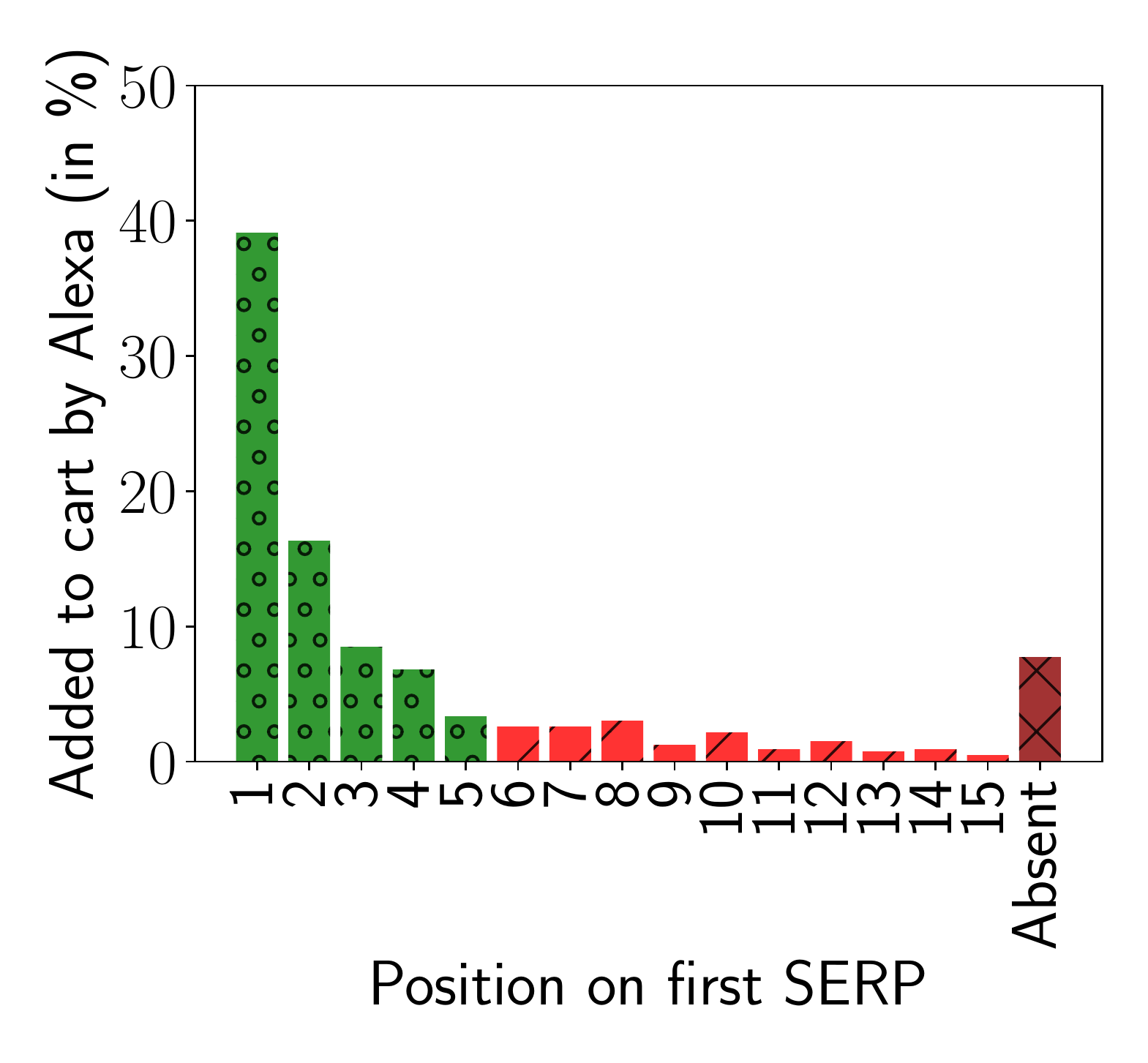}
		\vspace{-4 mm}
		\caption{Amazon's Choice -- Position on SERP}
	\end{subfigure}
	\begin{subfigure}{0.6\columnwidth}
		\centering
		\includegraphics[width= \textwidth, height=3cm]{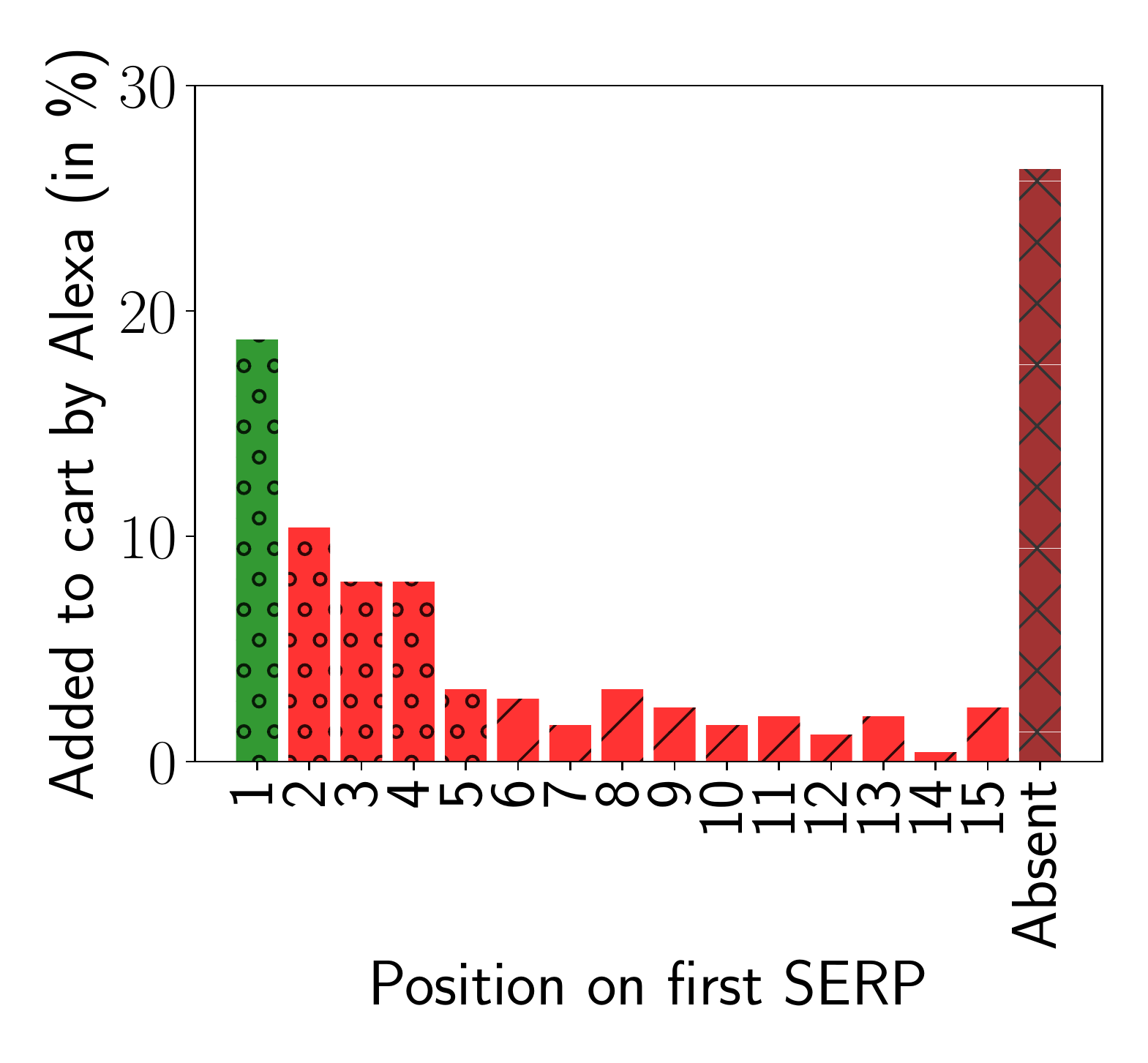}
		\vspace{-4 mm}
		\caption{A top result -- Position on SERP}
	\end{subfigure}
	\vspace{-4 mm}
	\caption{Break-up (in percentage) of rank of products that were added to cart by Alexa with different explanations, as per their (a)~price mentioned and (b, and c)~position on first SERP of Amazon desktop search. 
		Green (respectively, red) bars indicate positions where interpretations and observations match (respectively, do not match). 
		In a significant number of cases, a better option (as per Amazon's own desktop search) was available than the product that was added to cart by Alexa. 
	}
	\label{Fig: AC-Interpretation}
	\vspace{-5 mm}
\end{figure*}

\subsection{Interpretation of `Amazon's Choice'}
In Alexa's response to 66\% of the queries, we find `Amazon's Choice' to be the explanation for adding a product to the cart (e.g., Figure~\ref{Fig: Alexa-Transcripts}(a)). 
Amazon says that \textit{``Amazon's Choice highlights highly rated, well-priced products available to ship immediately''}~\cite{Amazon2021Choice}. 
To understand customers' interpretation of nuances like `highly rated', `well priced' etc., we asked the respondents the following questions: 

\noindent
$\bullet$ What do you interpret by a product to be ``highly rated''?

\noindent
$\bullet$ What do you interpret by a product to be ``well priced''?

\noindent
$\bullet$ If a product is explained to be ``Amazon's Choice'' for a query, where do you expect
that product to appear on your search results?

\noindent
$\bullet$ How likely are you to buy the product which is explained as ``Amazon's Choice'' for a query on Amazon?

Aggregated break-ups of responses to the questions related to Amazon's Choice explanation are shown in Figure~\ref{Fig: Survey-Interpretation-Responses}~(a--c) and (e). 

\vspace{1mm}
\noindent
\textbf{Interpretation of `highly rated': }
A total of 59\% of respondents voted that a highly rated product should be one which has an average user rating of greater than 4.0 (out of 5.0).
The two most voted options were the product should have an avg. user rating greater than or equal to 4.0 and 4.5 (33\% and 26\% votes respectively -- see Figure~\ref{Fig: Survey-Interpretation-Responses}(a)). Notice that many respondents (22\%) also interpreted this statement to be the best rated product among all the results shown on the SERP. Going with the plurality, we interpret a `highly rated' product to be a \textit{product having average user rating $\ge 4.0$.}

\noindent
\textbf{Observation in the collected data: }Out of the 662 queries for which a product was added to cart for being Amazon's choice, we observe that all 662 times the selected product has an average user rating $\ge$ 4.0 (out of 5). In fact, in nearly 16\% cases the product is rated higher than 4.5 too. Thus, \textit{Amazon abides by its claim that the Amazon's choice products are highly rated products} and the observation matches the interpretations of the respondents.

\vspace{1mm}
\noindent
\textbf{Interpretation of `well priced': }Interpretation of `well priced' product also seemed to be rather straight forward among the respondents. They consider a product to be well priced, if its price is among the least 5 prices among all the products shown in the search results for a query (61\% votes for the response-- Figure~\ref{Fig: Survey-Interpretation-Responses}(b)). 19\% of the respondents even went ahead to say that it should be the least priced product among all the relevant results shown.

\noindent
\textbf{Observation in the collected data: }To understand the interpretation of `well priced' product, we ranked all the products appearing in the first desktop SERP (for a particular query) as per the mentioned price. 
Notice in case of a tie between the price of two products, their position on the SERP was used to resolve the tie (such that every product will have a distinct position). 
Figure~\ref{Fig: AC-Interpretation}(a) shows the break-up of positions of different products that were added to cart by Alexa with Amazon's choice explanation (for different queries) based on their price.\footnote{Note that figures in Figure~\ref{Fig: AC-Interpretation} are truncated at position 15 for better visibility.} 
We observe that in merely 23\% cases, the product added to cart adhered to the most common interpretation of well-priced product as mentioned above. 
In other words, for as many as 77\% of the queries, the product selected by Alexa does not appear within the 5 least-priced offers available for the said query (contrary to what is understood by a majority of customers). 

\vspace{1 mm}
\noindent
\textbf{Expected position of appearance of Amazon's Choice: }Given an Amazon's choice product is highly rated and well priced, majority of the respondents either expect such a product to appear as the top search result (30\% respondents) or to at least appear among the top-5 products (54\% respondents) in the search results (Figure~\ref{Fig: Survey-Interpretation-Responses}(c)). 

\noindent 
\textbf{Observation in the collected data: }Figure~\ref{Fig: AC-Interpretation}(b) shows the break-up of positions of different products that were added to cart by Alexa with Amazon's choice explanation (for different queries) on the first desktop SERP.\footnote{Amazon search results 
	may have sponsored advertisements too. In our analyses, we consider only the organic results and not the sponsored ads.}. 
We observe that in 74\% cases (i.e., for 490 queries) Amazon's choice product was selected from the top-5 desktop search results. Contribution from position 1, however, is merely 39\%. 
In other words, {\it for 61\% of the cases (i.e., for 403 queries) there existed at least one or more products which the Amazon search system itself evaluated to be more relevant for the corresponding query at that point of time, yet which was not selected by Alexa.}

Even though the observation is in agreement with customers' expectations in 74\% cases, there is still a significant number of queries (172 out of 662 queries, i.e., 26\%) where a product positioned at or beyond rank 6 was added to cart with an explanation of being Amazon's choice (contrary to customers' expectations). 
Further, \textit{for nearly 8\% of the times (for 51 queries), the Amazon's choice product did not even appear on the first SERP} (the `Absent' bar in Figure~\ref{Fig: AC-Interpretation}(b)). 

\subsection{Interpretation of `a top result'}

For a significant number of queries (251) in our data collection process, the product was added to cart with `a top result' explanation (e.g., Figure~\ref{Fig: Alexa-Transcripts}(b)). To understand what the customers interpret being `a top result' we asked them the following questions:

\noindent
$\bullet$If a product is explained to be ``a top result'' for a query, where do you expect
that product to appear on your search results?

\noindent
$\bullet$ How likely are you to buy the product which is explained as ``a top result'' for a
query on Amazon?
Aggregated break-ups of responses to these questions 
are shown in Figure~\ref{Fig: Survey-Interpretation-Responses}~(d) and (f). 

\noindent {\bf Expected position of appearance of a top result:}
62\% of the respondents interpret a `top result' to be the top product in the search result; while another 33\% interpret a top result should be a product appearing in one of the top-5 positions in the search result (Figure~\ref{Fig: Survey-Interpretation-Responses} (e)). 
For the rest of this paper, we proceed with the interpretation having the majority of the votes, i.e., `a top result' means \textit{the top result} (position 1) in the search results. 

\noindent
\textbf{Observation in the collected data: }Figure~\ref{Fig: AC-Interpretation}(c) shows the break ups (in \%) of ranks (on the first SERP) from which different products were added to cart by Alexa system with `a top result' explanation. 
Contrary to the interpretation mentioned above, only 18.72\% of such products actually were positioned at the top of desktop search results (corresponding to position 1 in the figure) i.e. for \textbf{$\approx$81\%} of the cases the most popular interpretation does \textit{not} match with our observation. 
Even if we consider the second most popular interpretation of top-5, that leaves out nearly 52\% products which were ranked at position six and beyond. 
\textit{Worryingly, for 66 out of the 251 queries, the mentioned product (whose addition was explained by `a top result') did not even appear on the first desktop SERP page which was collected immediately after the query was posed to Alexa.}

We also performed similar ranking based analyses 
on the remaining 87 queries where explanations other than `Amazon's Choice' and `a top result' were given by Alexa. 
We found that across all 1000 queries, in only 32\% cases, the most relevant product (top-ranked product) according to desktop search was added to cart by Alexa. 
The details 
are added in the supplementary material (Figure~\ref{Fig: Likelihood-1000-Others}). 

\subsection{Implications from the survey }
In this section, we investigated the alignment of customers' interpretation of the two most prevalent explanations given by Alexa. 
A summary of the findings is noted in Table~\ref{Tab: Summary-Takeaways}. 
We observe that for `Amazon's Choice' and `a top result', the interpretation of the respondents and the observations from an immediate desktop search result do {\it not} conform with each other in several aspects. 
While Amazon's choice products are indeed highly rated, they are not what customers perceive to be `well priced'. The interpretation of `a top result' explanation is severely 
misunderstood by the customers.

Note that these observations are not based on a solitary snapshot of the search results. Our comprehensive analysis across different temporal snapshots also highlight such gaps between customers' interpretation and our observations on data collected from immediate desktop searches. 
The reader can refer to Figure~\ref{Fig: Temporal-Interpretation} in Section~\ref{Sec: Sup-temporal} of the supplementary material for further details.

Additionally, in the survey we had also asked our respondents about their likelihood to buy the products that are explained by Alexa to be `Amazon's choice' or `a top result'. 
Figure~\ref{Fig: Survey-Interpretation-Responses}(e) and Figure~\ref{Fig: Survey-Interpretation-Responses}(f) respectively show the responses obtained for ``Amazon's Choice'' and `a top results' explanations. In both cases, majority of the respondents (56\%) answered that they are `likely' or `very likely' to buy products with such explanations. 
This observation further emphasizes that such explanations act as powerful positive nudges 
for the customers. 
However, if the explanations do \textit{not} match their interpretations, then customers may be misled to products which they would not have purchased otherwise. 
It may not only result in customer dissatisfaction, but also result in decline in trust on the explanations (and therefore the response) of VAs such as Alexa.

\begin{table}[tb]
	\noindent
	\scriptsize
	\centering
	\begin{tabular}{ |p{1.7cm}|p{1.8 cm}|p{2.2cm}| |p{1.5cm}|}
		\hline
		{\bf Explanation type } & {\bf Statement} & {\bf Interpretation} & Match\\
		\hline
		& Highly rated & Avg. user rating $\ge$ 4.0 & \cmark (100\%) \xmark(00\%) \\
		\cline{2-4}
		{\bf Amazon's Choice }& Well priced & Least-5 price & \cmark (23\%) \xmark(77\%) \\
		\cline{2-4}
		& Expected position & Top-5 in SERP & \cmark (74\%) \xmark(26\%)\\
		\hline
		{\bf A top result} & Expected position & Top result (position 1)& \cmark (19\%) \xmark(81\%) \\
		\hline
	\end{tabular}	
	\caption{{\bf Major takeaways from Section~\ref{Sec: Interpretation}. While Amazon's choice products are 
	highly rated; there is a significant mismatch in the interpretation of 
	a `well-priced' product. 
	The interpretation of `a top result' is severely misinterpreted too
	.}}
	\label{Tab: Summary-Takeaways}
	\vspace{-10 mm}
\end{table}
	\section{(Un)Fairness in product selection}
\label{Sec: Deservingness}
While traditional (desktop) e-commerce search shows a ranked list of products
, a voice assistant, in contrast, selects only one product and adds it to the customers' cart for further exploration and purchase (see Figure~\ref{Fig: Search-Mediums}). 
Note that here we are considering the same user-account issuing the same query at the same time instant from the same geographical location with the same delivery location and, thereby, making the context of the VA and desktop searches as similar as possible. 
Now, if the product being added to the cart (by the VA) does {\it not} match with the most relevant product as per desktop search, there may arise a case for unfairness. 

\noindent
\textbf{(Un)fairness concerns: }
A product being added to cart significantly boosts its (and consequently its producer's) opportunity for sales. In addition, explanations such as those analysed in the previous section, may reinforce the likelihood of purchase among customers.
Considering the limited autonomy of customers in voice search, selection of a less relevant product may have unfair consequences for both producers and customers.
In spite of being the producer of the most relevant product (as per desktop search), \textit{one will be denied of the opportunity to sales due to non-selection by the VA}. 
Again, even though there exist one or more better products, \textit{a customer may end up purchasing a product that is (possibly) not up to the mark due to non-selection of the most relevant option.} 

In the present context, we have already shown that a significant majority of the products added to cart by Alexa do not belong to position 1 in the corresponding Amazon desktop search results (see Figure~\ref{Fig: AC-Interpretation}(b) and Figure~\ref{Fig: AC-Interpretation}(c)). 
Only 39\% and 18.72\% of the products added to cart with `Amazon's Choice' and `a top result' explanations respectively are from position 1. 
The percentage is around 32\% out of all 1000 queries. 
In other words, Amazon's own search ranking system evaluates that in 68\% cases 
there exists at least one product which is more relevant (to the same query, and in exactly the same setting) than the product added to cart by Alexa.

In the remainder of this section, we will quantify and investigate unfairness and bias (if any) in the decisions taken by Alexa, from the perspective of the two major stakeholders in the e-commerce setup, i.e., producers, and customers. 

\subsection{(Un)Fairness toward producers} 
As mentioned in Section~\ref{Sec: Data}, along with the Alexa responses and the product added to cart, we also simultaneously collected the search results on desktop version. 
Now, we quantify the difference between the exposure that the product added to cart by Alexa (and thereby its producer/seller) gets due to its selection, and the exposure it would have got due to its placement in the desktop search results.

\noindent
\subsubsection{\textbf{Exposure due to Alexa search: }}Provided that Alexa VA used in the study adds a specific product to the cart for further exploration and/or purchase for a specific query, the exposure of the corresponding product (and its producer) is 1. For all the rest of the products, the exposure is 0. Mathematically, for an item $i$ and a query $q$, the exposure due to the Alexa VA can be operationalised as, $E_{Alexa}(i) = 1$, \textit{if i is in cart for query q} and 0 otherwise.

\noindent
\subsubsection{\textbf{Exposure due to desktop search:}} Amazon desktop search usually provides a ranked list of items sorted in decreasing order or relevance. 
To evaluate the exposure of a product due to desktop search, we assume that attention of the positions (and the exposure thereof) are distributed geometrically with a parameter $p$ (which is the probability of a search result being clicked) up to the position $k$. 
Note that geometrically distributed weights is a special case of the cascade model used in multiple prior works~\cite{biega2018equity, craswell2008experimental, dash2021when}. 
Mathematically, provided an ordered list of items $R$, with items ranked from $1$ to $k$, the exposure of a product $i$ at rank $r$ is operationalised as $E_{Desktop}(i) = p (1-p)^{r-1}$ (see~\cite{biega2018equity, craswell2008experimental, dash2021when} for details).
However, for the purpose of our analyses, we do not require the exposure of all items in the ranked list. Rather, we only require the exposure of the product which had been added to cart by Alexa. Note that the maximum exposure of a product at the top of the search result is $p$ as per the above formulation. Hence, for the evaluation of exposure bias of a single product (the one added to cart), we normalize these values with respect to $p$, i.e., $E_{Desktop}(i) = (1-p)^{r-1}$. 
We consider $p=0.35$ (the probability of a product being clicked) 
because 35\% of shoppers click on the first product 
on an Amazon SERP~\cite{Rubin2021Cracking}.

\noindent
\subsubsection{\textbf{Exposure bias due to selection by Alexa: }} Our aim here is to quantify the difference between a product's exposure due to its selection by Alexa and the exposure it would have got by virtue of its position in a ranked list; we call this difference {\it exposure bias}.

In this work, we consider three ranked lists as baselines for evaluation of the exposure bias --  ranked lists (1)~based on overall relevance to the query, as obtained from Amazon desktop search results, (2)~based on price of the products on the first SERP, and (3)~based on the average user ratings of the products on the first SERP.\footnote{For the latter two ranked lists, if there is a tie between two items, the tie is decided by their position as per Amazon search result, i.e., their relevance.}
The exposure of a product (and its producer) added to cart by Alexa is compared against its exposure in each of the different ranked lists for all queries. 
Given the Alexa and desktop exposure distributions of a query-set $Q$, we quantify exposure bias ($bias_{Exposure}$) as the distance between the two distributions. Note that a variety of measures can be applied here, e.g., KL divergence or L1 distance. In this paper, we measure $bias_{Exposure}$ using the latter, i.e.,: $bias_{Exposure} = \sum_{q\in Q} {|E_{Alexa}(cart(q))-E_{Desktop}(cart(q))|}$ 
where $cart(q)$ is the item added to cart by Alexa for the query $q$. Note that $bias_{Exposure}$ is normalized between $[0, 1]$, with $0$ denoting no exposure bias and $1$ denoting maximum exposure bias. 

\begin{table}[tb]
	\noindent
	\small
	\centering
	\begin{tabular}{ |p {3.0 cm}| p {1.2 cm}|p {0.8 cm}|p {0.8 cm}||p {0.8 cm}|}
		\hline
		\bf Ground-truth &\bf Amazon's Choice & \bf Top result & \bf Others & \bf Overall\\
		\hline
		Based on position on SERP & 0.43 & 0.68 & 0.74 & 0.52\\
		\hline
		Based on price on SERP & 0.88 & 0.86 & 0.86 & 0.87\\ 
		\hline
		Based on rating on SERP & 0.79 & 0.81 & 0.77 & 0.79\\
		\hline
	\end{tabular}	
	\caption{Mean bias scores due to product selection by Alexa, based on the different baseline rankings and the mentioned explanation type. Each value is the mean over all queries for which a particular explanation is given. 
	Higher the values, more unfair is the selection by Alexa.
	}
	\label{Tab: BiasBreakUp}
	\vspace{-10 mm}
\end{table}

\begin{figure}[t]
	\centering
	\begin{subfigure}{0.48\columnwidth}
		\centering
		\includegraphics[width= \textwidth, height=3cm]{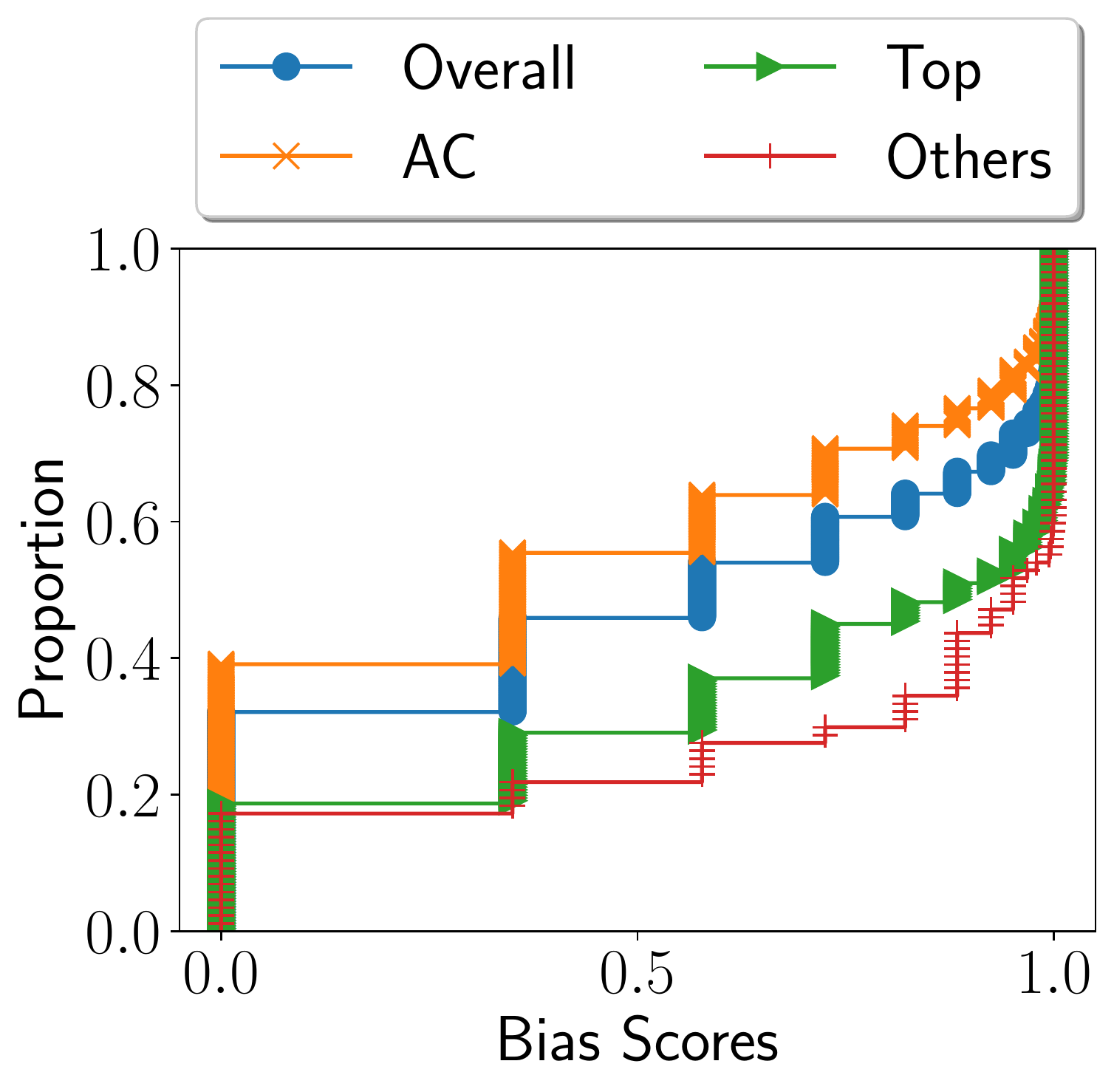}
		\vspace{-6 mm}
		\caption{}
	\end{subfigure}
	\begin{subfigure}{0.48\columnwidth}
		\centering
		\includegraphics[width= \textwidth, height=3cm]{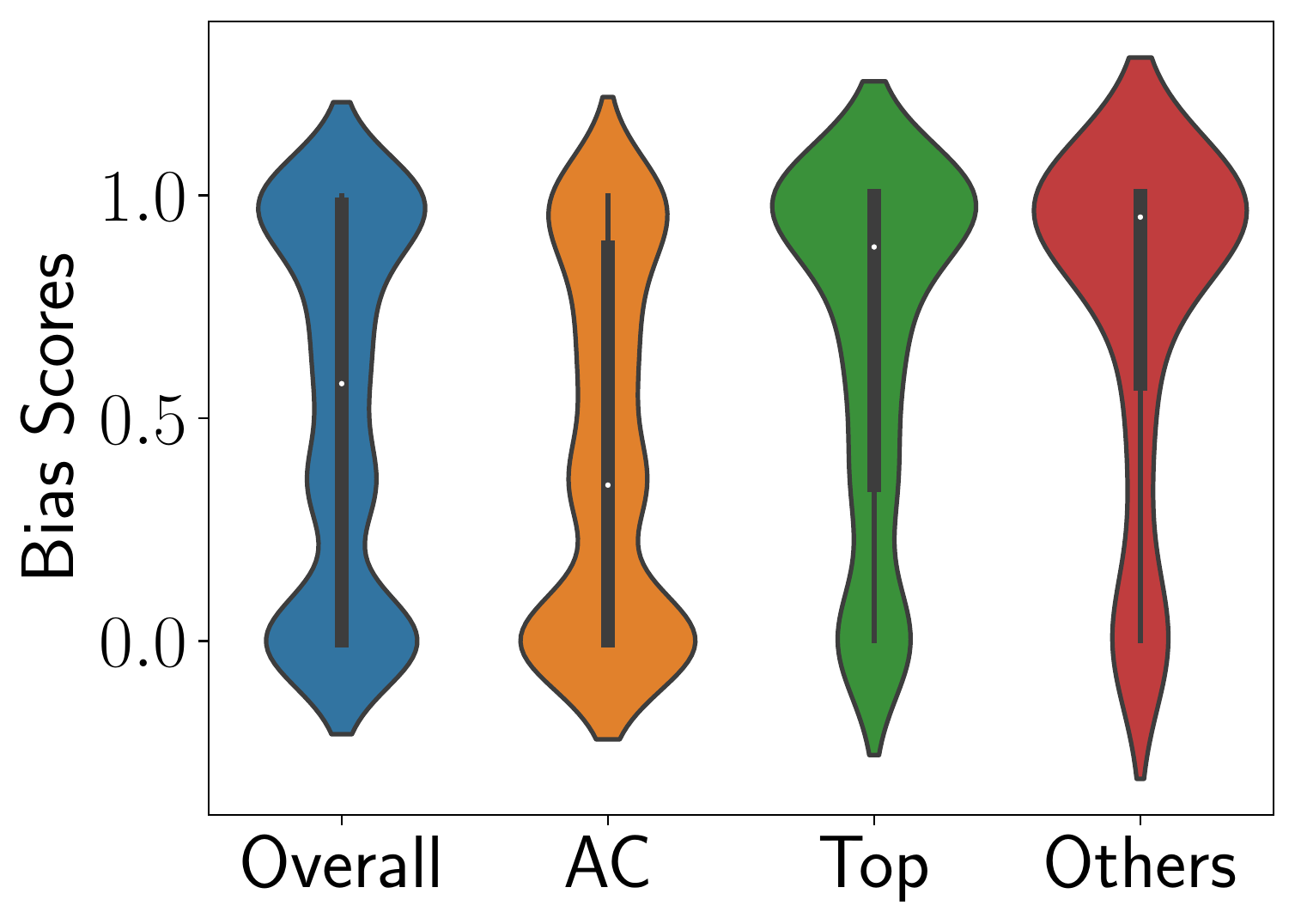}
		\vspace{-6 mm}
		\caption{}
	\end{subfigure}
	\vspace{-4 mm}
	\caption{ (a)~CDF, and (b)~violin plot of bias score distributions with position on SERP as the baseline segregated by explanation types. AC: Amazon's Choice, Top: `a top result'. Overall for only 32\% queries the bias score was 0, i.e., the most relevant product was added to cart. 
	The high width toward bias score 1 for the violin plot with `a top result' explanation suggests that even though the explanation was top result, the selected products came from lower positions.
	}
	\label{Fig: Bias-Score-Distribution}
	\vspace{-6 mm}
\end{figure}

\subsubsection{\textbf{Observations:} }
The mean $bias_{Exposure}$ with different baseline  ranked lists for different explanations is shown in Table~\ref{Tab: BiasBreakUp}. The last column (Overall) shows the results aggregated over all the 1000 queries. The higher bias scores with respect to price and rating-based baselines further suggest that, 
though products with lower price and / or better user-ratings were available for majority of the queries at the top of the desktop search results, Alexa added a product with relatively poor rating and / or higher price for to the cart. 
Bias with respect to position on the Amazon SERP (based on relevance) is significantly lower than the other two baseline ranks. This suggests that the selected items were coming from top positions in some cases (corroborating our observations in Figure~\ref{Fig: AC-Interpretation}).

Figure~\ref{Fig: Bias-Score-Distribution} shows the CDF and violin plot of the bias score distribution with respect to the position on SERP as the baseline. 
We see (from the blue curve in Figure~\ref{Fig: Bias-Score-Distribution}~(a)) that for merely 32\% of all queries, the top desktop search result was actually added to cart by Alexa (leading to a bias score of 0). 
More alarmingly, this percentage drops drastically to below 20\% for explanations related to `a top result' and other explanations. 
For a significant fraction of queries, the score was close to 1 (very high bias) as indicated by larger width of the blue violin plot for overall cases (Figure~\ref{Fig: Bias-Score-Distribution}~(b)). 
This is caused due to the inconsistency of `a top result' explanation, as discussed earlier. Even though the explanation says it is a top result, the product selected into the cart does {\it not} necessarily align with this explanation in majority of the cases. 
Distributions of other baselines indicate similar trends too
 (see Figure~\ref{Fig: Ground-Truth-1000} in the supplementary material). These observations further highlight the apprehended unfairness concerns toward producers 
due to non-selection (by Alexa) of most relevant products.

\subsection{(Un)Fairness toward customers}~\label{Sec: SurveyPreference}
Even with high values of exposure bias scores w.r.t. different baseline rankings, one can argue that if the Alexa-selected products are preferable to the customers, then at least from the customer's point of view this situation may be acceptable. 
To ascertain if this is the case, we conducted another survey with the same 100 participants about their preference between products. 

\vspace{1 mm}
\noindent
\textbf{Survey setup: }
The participants were asked to choose between a pair of products for a given query -- (i)~the product which was added to cart by Alexa for the said query, and (ii)~the top-ranked result of the Amazon desktop search ranking for the same query and at the very same instant. 
We went ahead with product (ii) for the comparison since, according to Amazon's own search system, it is the most relevant product for the query at the time for the corresponding customer. 
During the survey we showed a participant -- the title of the two products, their prices, their average user ratings and their number of ratings received by each of the two products (as shown in the Amazon SERP during our data collection). 
Given the two products as options, we asked a participant the following two questions: (1)~\textit{Suppose you are looking for "<query string>". Which of the following would you prefer to buy?}, and 
and (2)~\textit{Briefly explain your selection} -- to understand the reasons behind their preference.
Overall, we evaluated 30 distinct queries for which the Alexa-selected product was different from the top desktop result. Each participant responded with their preference to 10 different queries.

\vspace{1 mm}
\noindent
\textbf{Observations:} Out of the 1000 evaluations (100 respondents $\times$ 10 queries), \textbf{\textit{732 (i.e., 73.2\%)} times the participants chose the top desktop search result over the product added to cart by Alexa} for the corresponding queries. 
This observation indicates the overwhelming rejection of participants for products selected by Alexa, thus underpinning (un)fairness concerns discussed earlier. 

\begin{figure}[t]
	\centering
	\begin{subfigure}{0.45\columnwidth}
		\centering
		\includegraphics[width= \textwidth, height=2.4cm]{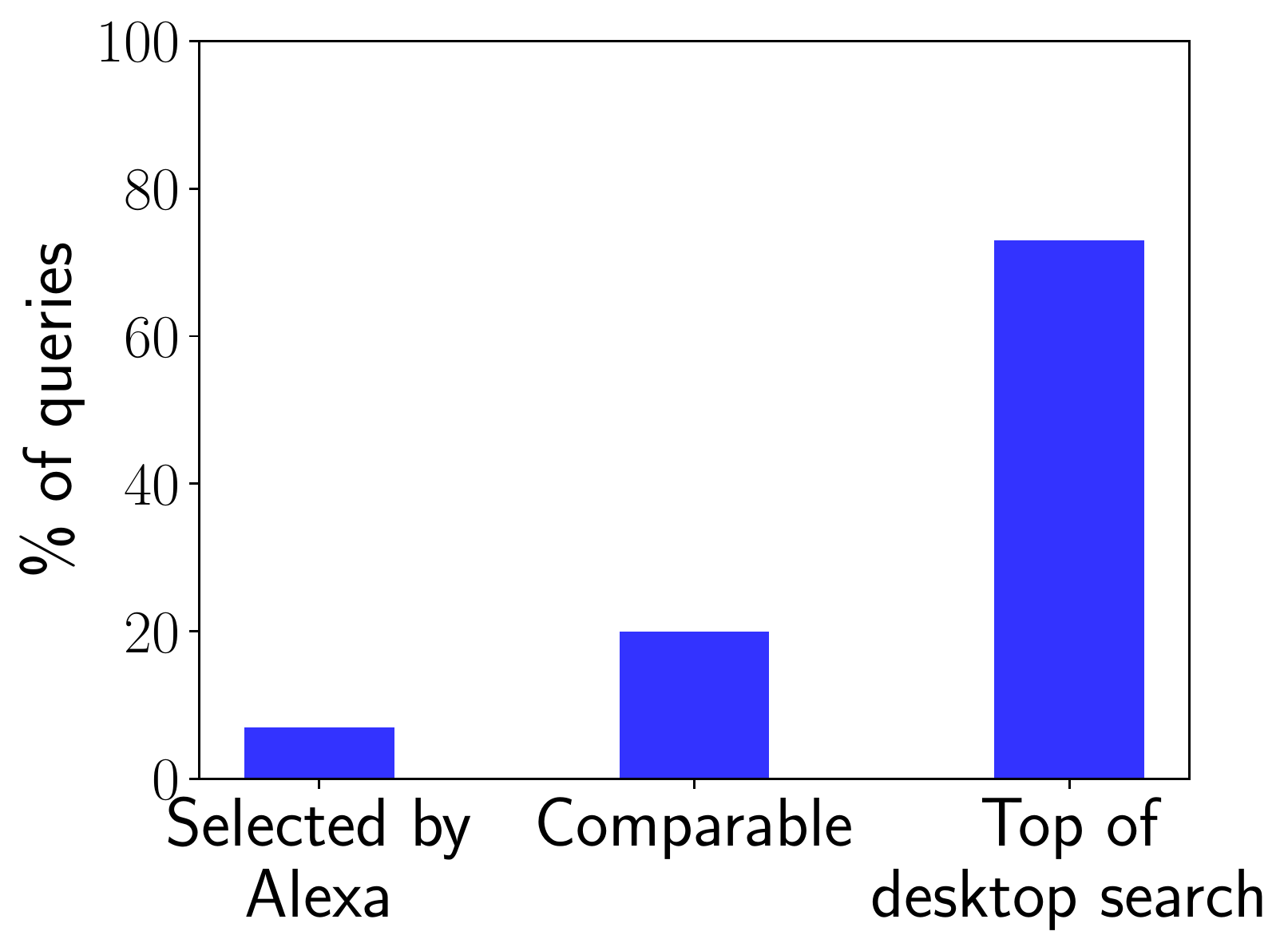}
		\vspace{-6 mm}
		\caption{}
	\end{subfigure}
	\begin{subfigure}{0.49\columnwidth}
		\centering
		\includegraphics[width= \textwidth, height=2.4cm]{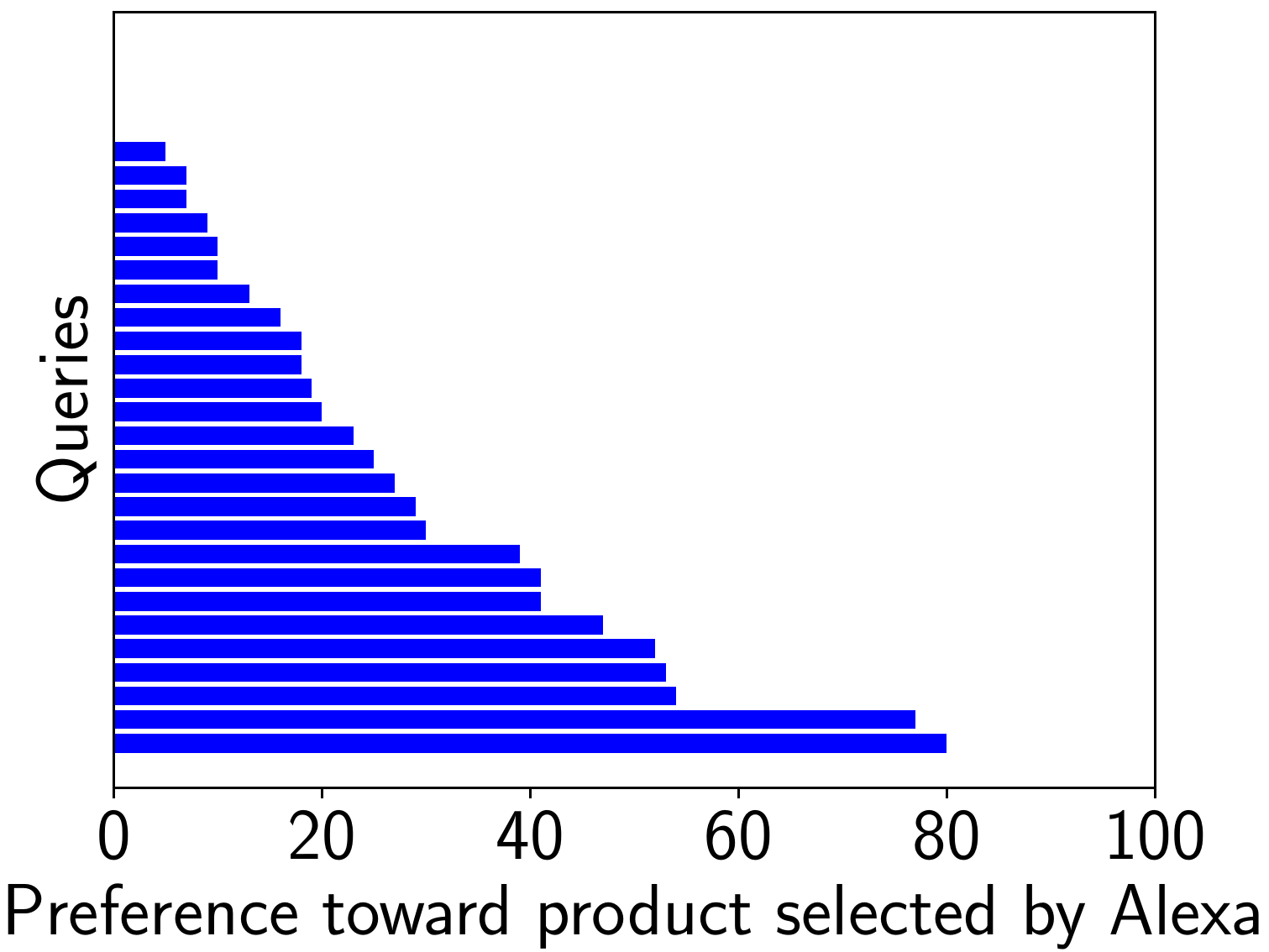}.
		\vspace{-6 mm}
		\caption{}
	\end{subfigure}
	\vspace{-4 mm}
	\caption{(a)~The \% of queries for which preference of the survey participants matches with that by Alexa. 
		(b)~The \% of participants who preferred the Alexa-selected product. For 22 out of 30 distinct queries, participants preferred the top desktop search result to the Alexa-selected product.}
	\label{Fig: Survey-Query-Results}
	\vspace{-7 mm}
\end{figure}

Figure~\ref{Fig: Survey-Query-Results} shows the results for different queries for which participants were asked to select between two products. 
We consider a margin of 20\% gap to be a significant majority. In other words, if 60\% or more respondents vote for one product over the other, we consider that product to be the overwhelming majority for the corresponding query. 
Out of the 30 distinct queries for which we collected responses, the people's selection and Alexa's selection match with overwhelming majority for only 2 queries. 
For six queries (20\% cases), the responses were split between the Alexa-selected product and the top desktop search result with no clear preference. 
For the rest 73.33\% of the cases (i.e., 22 out of 30 queries), the participants preferred the top desktop search result with more than 60\% votes. 

\vspace{1 mm}
\noindent
\textbf{Trends from the participants' explanations:} We manually examined the explanations given by the participants for their slections (i.e., the responses provided to question~(2)); we observe the following key trends. 
Brand name matters -- for many of the queries, even though the competing product is highly rated, 
participants generally opted for more reputed brands. 
High average user-rating alone did not persuade many of the participants; high user-ratings along with higher number of reviews is considered to be more preferable. 
Also, when participants are provided with two equally rated products, or two products of the same brand, they often prefer the cheaper one. 

\noindent
\textbf{Takeaways: }
Adding a product to the cart as part of the status quo action is an explicit endorsement from the VA (and its choice architects) for the product. Therefore, the non-selection of most relevant (or best-priced or best-rated) products in such significant percentage 
of queries highlights serious unfairness concerns for the producers (and/or sellers) of those products.  Worryingly, we also find qualitatively similar observations across the temporal snapshots (see Table~\ref{Tab: BiasBreakUp-Temporal} and Figure~\ref{Fig: Temporal-Bias-Distributions} in the supplementary material).

From the customer's perspective, we observed that respondents mostly prefer to buy the products appearing at the top of desktop search results to the ones added to cart by Alexa.
This observation further highlights the unfairness issues toward customers and the resultant customer dissatisfaction that may arise due to the default product selection by Alexa.

	\section{Related works}


\vspace{1 mm }
\noindent
\textbf{Intelligent voice assistants: }
Several prior works have showed the impact of the voice and information quality of VAs having positive effect on consumer trust and further willingness to use these systems~\cite{nasirian2017ai, poushneh2021humanizing, foehr2020alexa}.
Security and privacy risks associated with VAs have also been investigated~\cite{chung2017alexa}, calling for better diagnostic testings to ensure more trustworthy VAs.
Several cognitive biases (e.g., priming and anchoring biases) during the interaction of VAs and customers have also been studied in prior works~\cite{kiesel2021meant, santhanam2020studying}.
While these prior works discuss about some important aspects
, none of them investigates the understanding of humans about the framing of different responses by VAs (which we do in this work). 

\vspace{1mm}
\noindent
\textbf{Bias and unfairness in information access systems: }
A rich vein of studies have focused on issues related to fairness of information access algorithms, ranging from individual fairness~\cite{biega2018equity, singh2019policy, lahoti2019ifair, patro2020fairrec, patro2020incremental} to group fairness~\cite{zehlike2017fa, geyik2019fairness, singh2018fairness, dash2019summarizing}. 
Cognitive biases due to nudges from information access system have been investigated in multiple studies as well~\cite{baeza2018bias, schneider2018digital, azzopardi2021cognitive, novin2017making, chakraborty2019impact, mota2020desiderata}.

The current work is a suitable amalgamation of studying interpretation (from cognitive viewpoint) and fairness issues (from the perspectives of producers, and customers 
) due to responses provided by Alexa systems upon different e-commerce search queries.

	\section{Concluding discussion}

To our knowledge, this is the first attempt to understand the implication of responses (explanation and default action) provided by voice assistants during e-commerce search. 
We observe significant mismatch between the human interpretation of different explanations from Alexa and the actual observation on Amazon desktop search. 
Through our user survey, we also observed that customers would often prefer to buy the products appearing as the top desktop search results, to the one that is added to cart by Alexa. 
These findings underline the importance of bridging the gap between the framing of a VA's responses and the interpretation by the customers. 
Since the amount of choices presented is significantly low, users tend to cede more autonomy to the decisions taken by the VAs. 
Thus, it is important that VAs lead customers to the most relevant 
products for their queries, adhering to their expectations. 

\vspace{1 mm}
\noindent
\textbf{Future directions:} This work may open up a number of research directions in future. 
Multiple media articles~\cite{Kalra2021AmazonOct, amazon-nytimes-creswell} and prior works~\cite{dash2021when} have introduced new sources of unfairness in e-commerce marketplaces due to special relationships between stakeholders (e.g., due to private label products). 
We intend to extend our investigation to such concerns in future.
Though the current work is focused only on e-commerce search, the research questions can be extended (with some variations) to any generic search or QA operations on VAs. 
Finally, while cognitive biases have been extensively studied in psychology~\cite{kahneman2011thinking, thaler2008nudge}, information access systems~\cite{azzopardi2021cognitive, novin2017making}, and conversational systems~\cite{santhanam2020studying} separately, information access through VAs provides a new paradigm of exploration such biases.


	\vspace{3mm}

	\begin{acks}
	This research is supported in part by a European Research Council (ERC) Advanced Grant for the project ``Foundations for Fair Social Computing", funded under the European Union's Horizon 2020 Framework Programme (grant agreement no. 789373), and by a grant from the Max Planck Society through a Max Planck Partner Group at IIT Kharagpur. A. Dash is supported by a fellowship from Tata Consultancy Services.
	\end{acks}

	\bibliographystyle{ACM-Reference-Format}
	\bibliography{Main}
	\newpage
	\section{Supplementary Material}

In this material, we provide some additional results for the 1000 query snapshot as well as results for the temporal snapshots collected for 100 queries over a period of two weeks. The break up of number of  products with different explanations added to cart by Alexa is shown in Table~\ref{Tab: BasicBreakUp}. For the temporal snapshots of 100 queries, the percentage of products with Aamzon's Choice explanation is 5\% more than that for the 1000 query snapshot.

\begin{table}[b]
	\noindent
	\small
	\centering
	\begin{tabular}{ |p {1.8 cm}| p {2.5 cm}|  p {1.5 cm}|p {1.1 cm}|}
		\hline
		\bf \# Queries &\bf \# Amazon's Choice  & \bf \# Top result & \bf \# Others\\
		\hline
		1000 & 662 (66.2\%) & 251 (25.1\%) & 87 (8.7\%)\\
		\hline
		100 $\times$ 14 = 1400 & 1002 (71.6\%) & 343 (24.5\%) & 55 (3.9\%)\\
		\hline
	\end{tabular}	
	\caption{{\bf Total number of queries and break up of what fraction of times products with Amazon's choice, or top results or other explanations were added to cart by Alexa.}}
	\label{Tab: BasicBreakUp}
	\vspace{-8mm}
\end{table}

\subsection{Additional results for 1000 query snapshot}
\begin{figure}[b]
	\centering
	\begin{subfigure}{0.48\columnwidth}
		\centering
		\includegraphics[width= \textwidth, height=3cm]{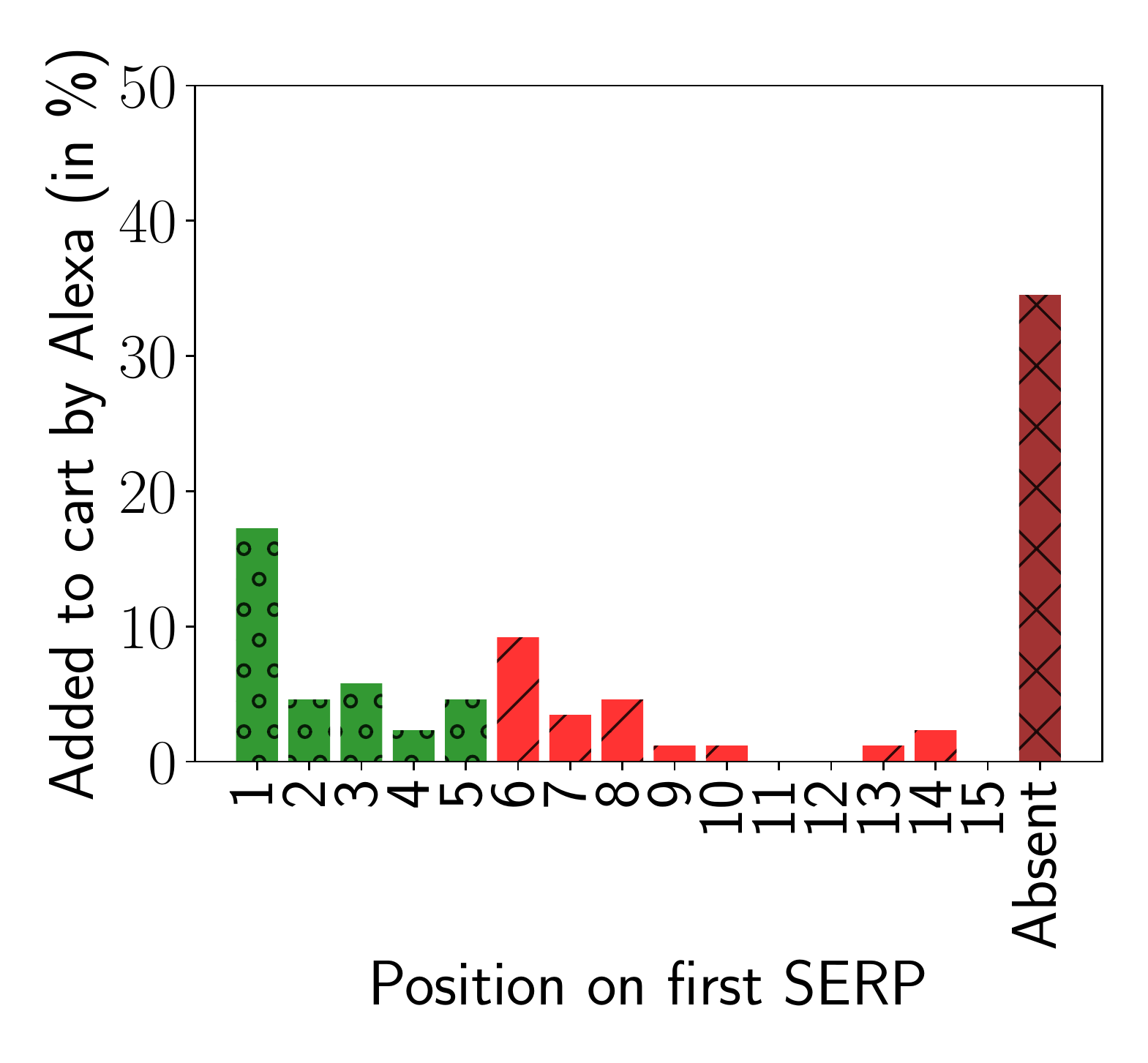}
		\caption{Other explanations}
	\end{subfigure}
	\begin{subfigure}{0.48\columnwidth}
		\centering
		\includegraphics[width= \textwidth, height=3cm]{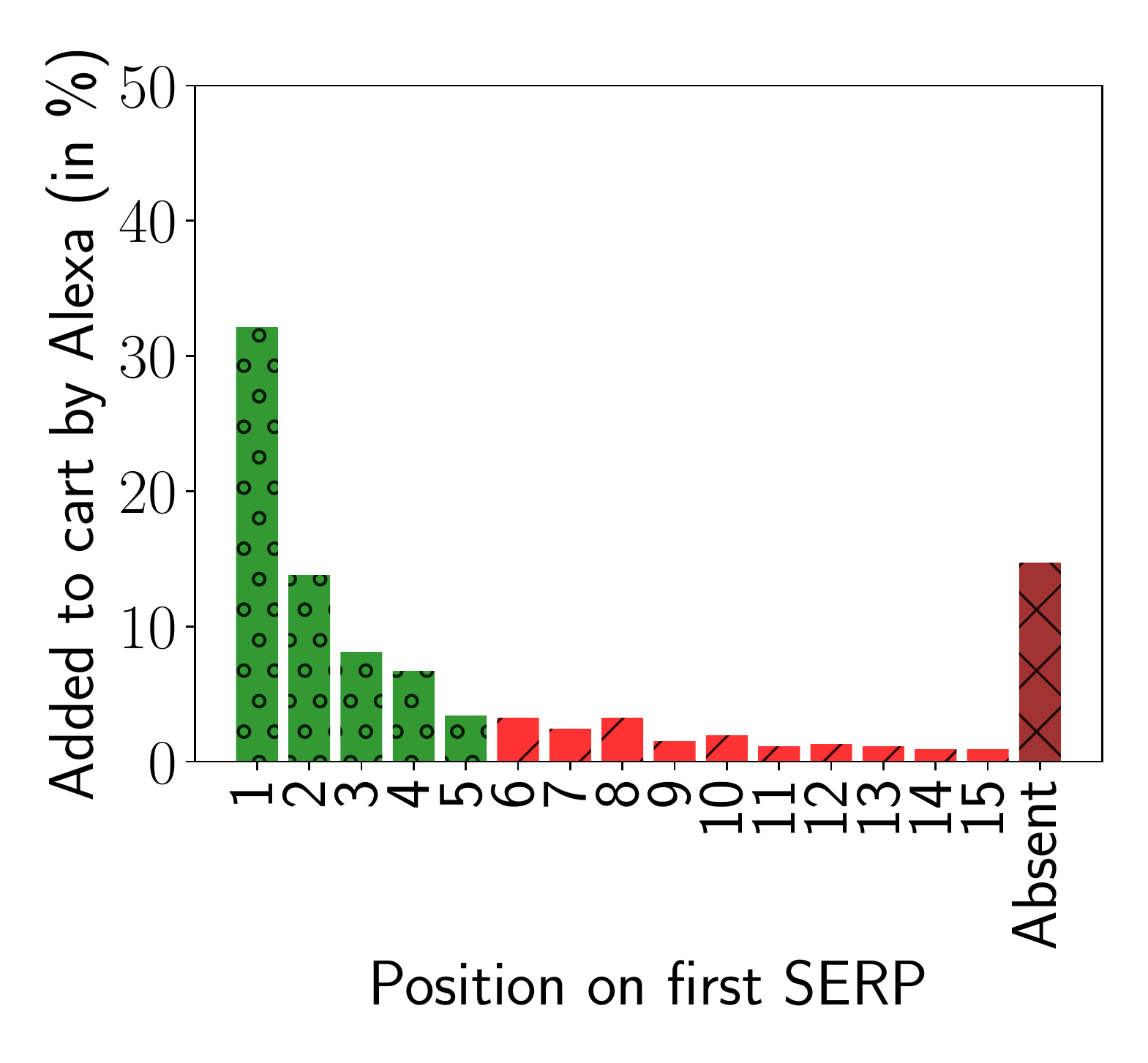}
		\caption{Overall 1000 queries}
	\end{subfigure}
	\caption{ The break ups of rank of different products, as per their position on first SERP, that were added to cart with (a)~explanations other than Amazon's Choice and `a top result', (b)~for all 1000 queries. The distributions suggest that in significant number of cases, a better option was available than the product which was added to cart by Alexa.
	}
	\label{Fig: Likelihood-1000-Others}
	\vspace{-5 mm}
\end{figure}

\textbf{Rank wise break ups for other explanations}: Much like in Figure~\ref{Fig: AC-Interpretation}(b--c), the distributions in Figure~\ref{Fig: Likelihood-1000-Others} also suggest that a significant fraction of products were added to cart from lower search result positions by Alexa. Figure~\ref{Fig: Likelihood-1000-Others}(b) suggests that overall only in 32\% cases the most relevant product (as per Amazon's own search result on desktop for the same user at the same time) was added to cart by Alexa. 

\begin{figure}[t]
	\centering
	\begin{subfigure}{0.48\columnwidth}
		\centering
		\includegraphics[width= \textwidth, height=3cm]{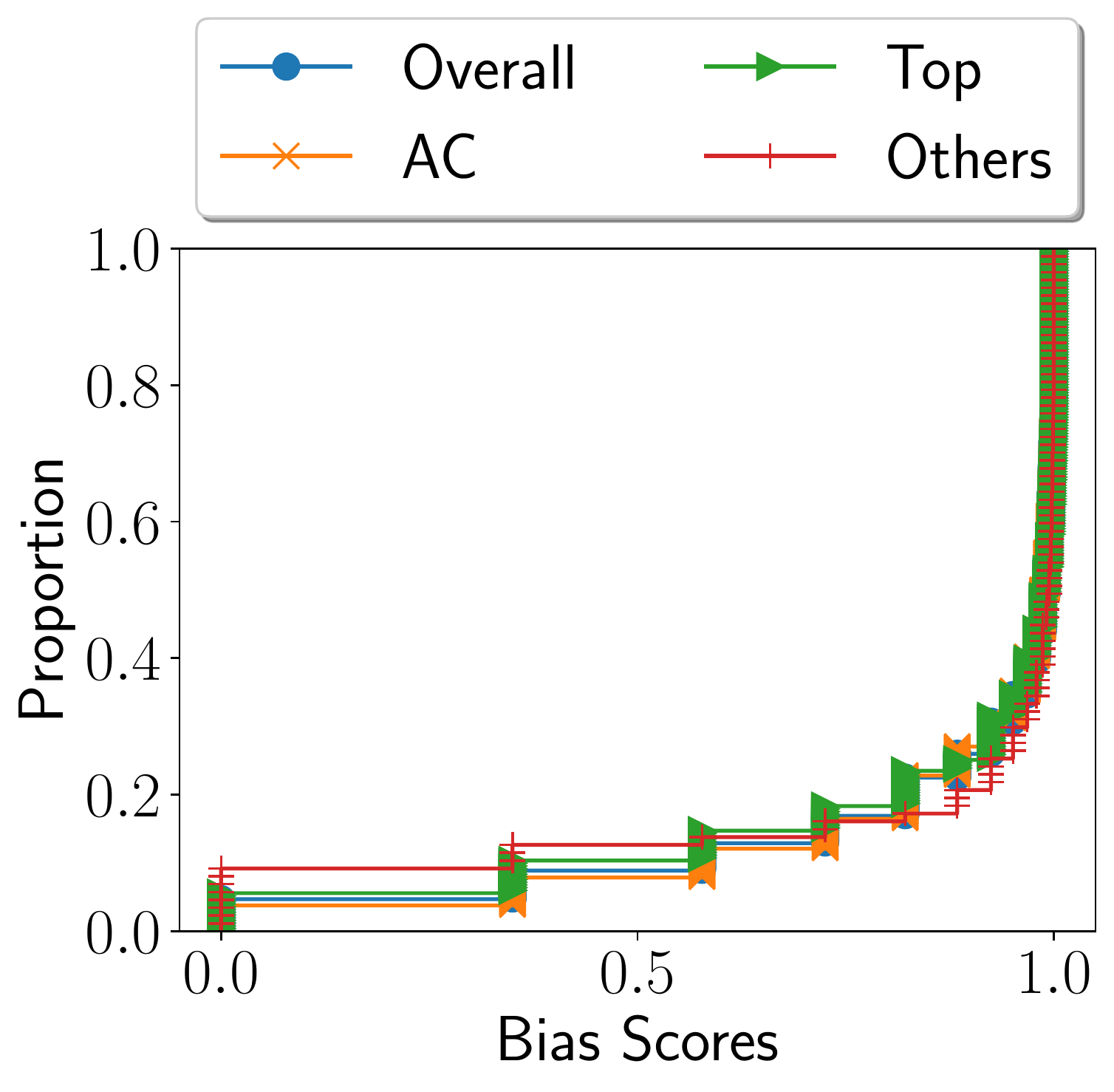}
		\caption{Other explanations}
	\end{subfigure}
	\begin{subfigure}{0.48\columnwidth}
		\centering
		\includegraphics[width= \textwidth, height=3cm]{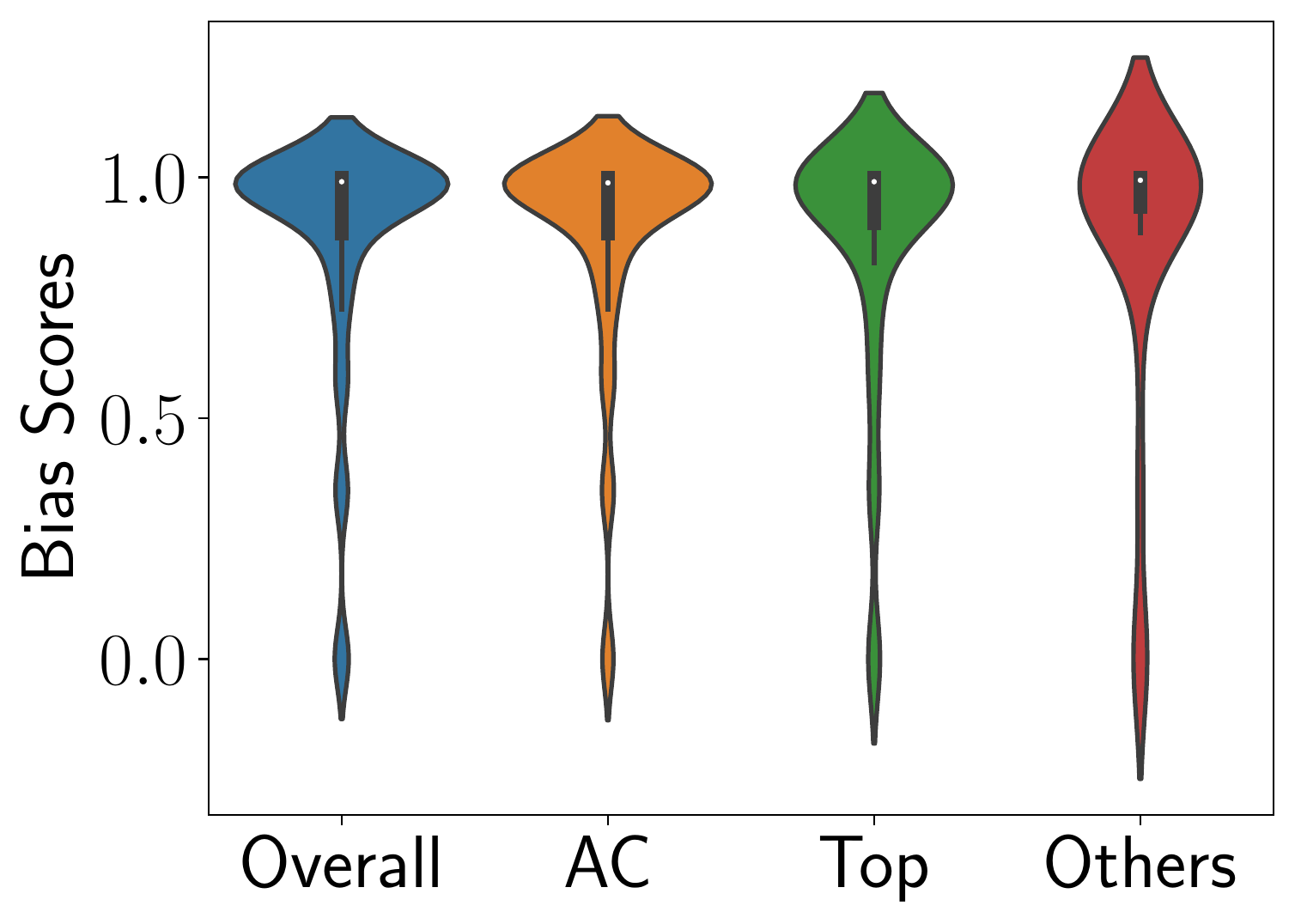}
		\caption{Overall 1000 queries}
	\end{subfigure}
	
	\begin{subfigure}{0.48\columnwidth}
		\centering
		\includegraphics[width= \textwidth, height=3cm]{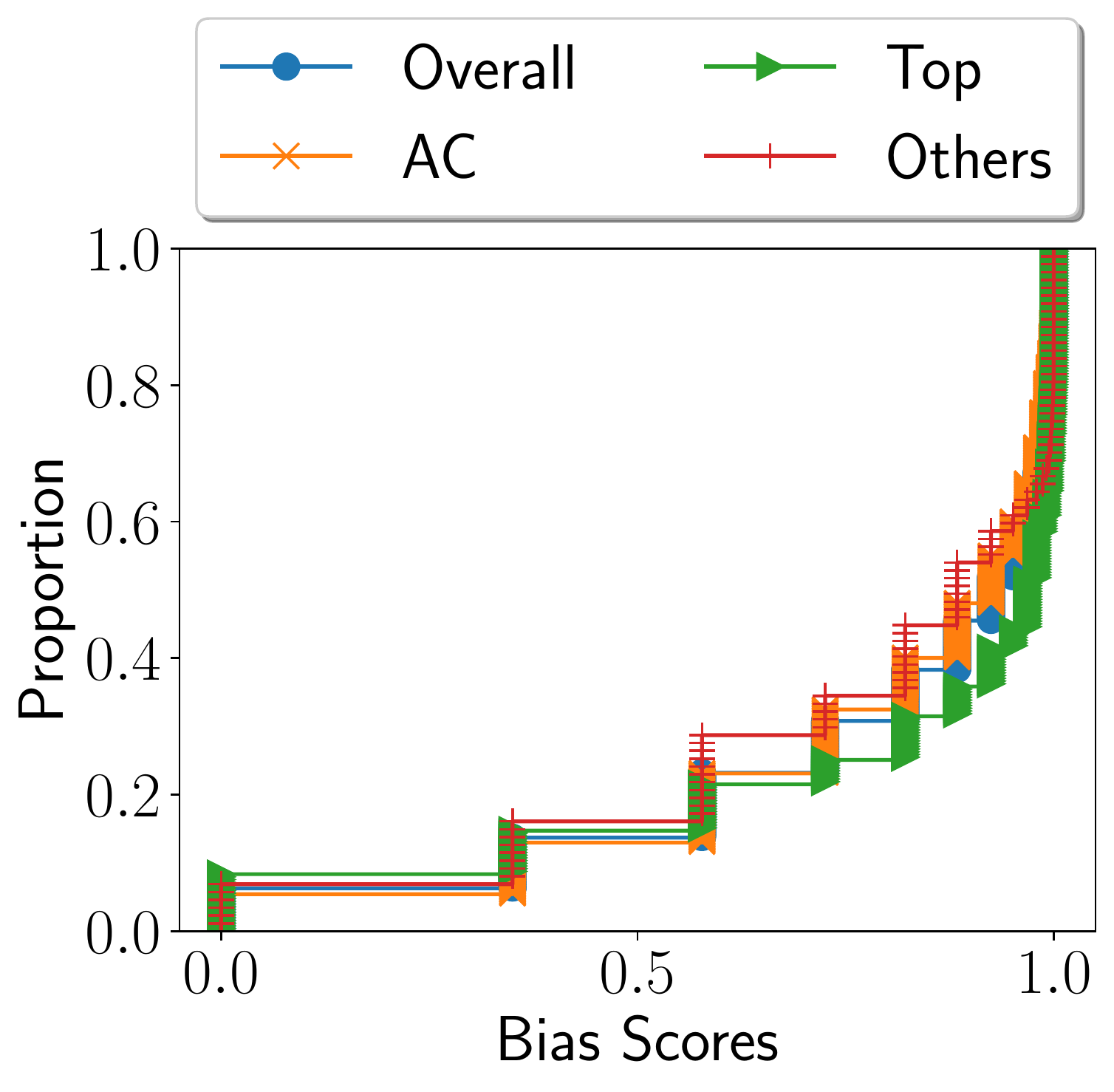}
		\caption{Other explanations}
	\end{subfigure}
	\begin{subfigure}{0.48\columnwidth}
		\centering
		\includegraphics[width= \textwidth, height=3cm]{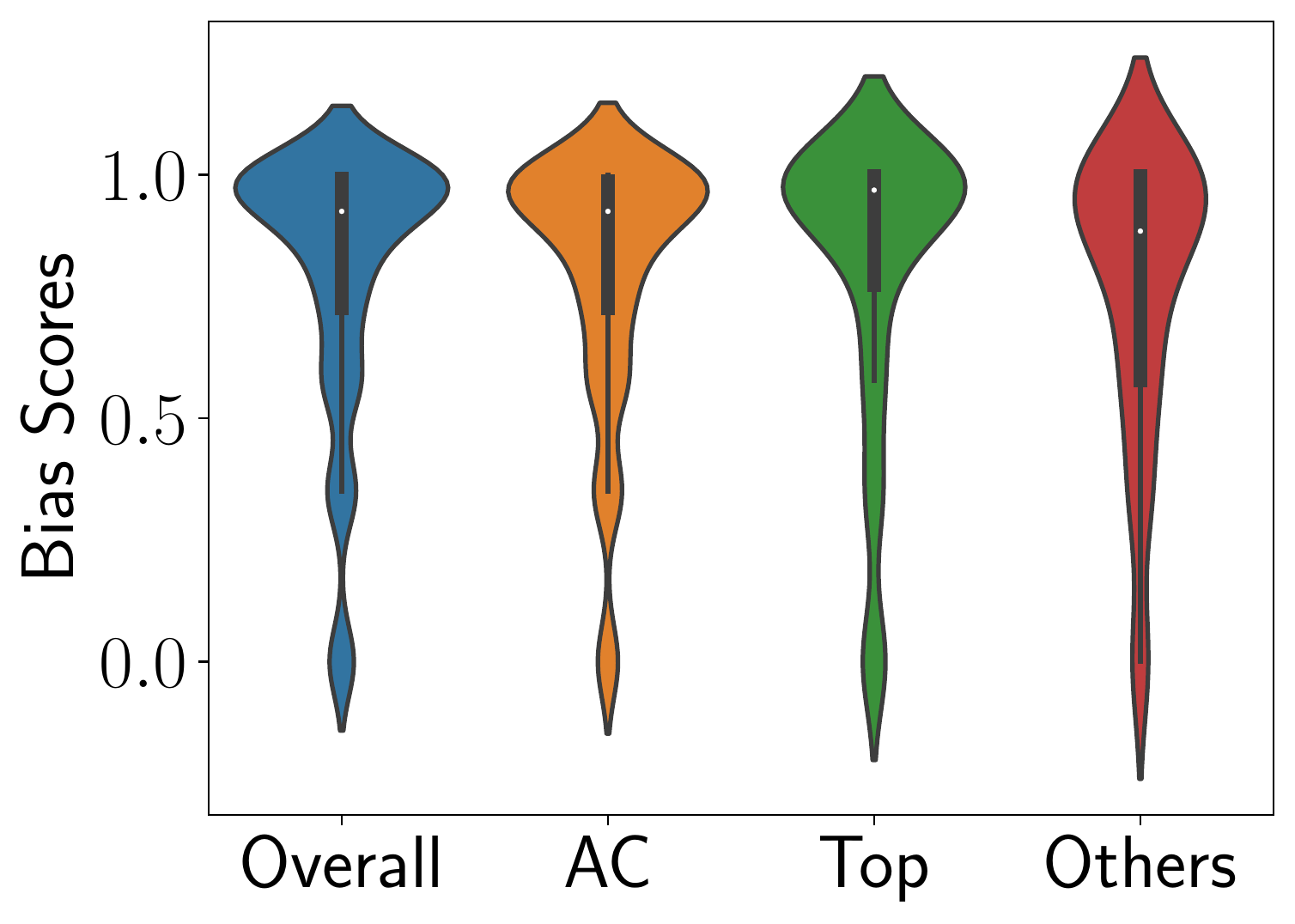}
		\caption{Overall 1000 queries}
	\end{subfigure}
	\caption{ CDF and violin plot of the bias score distributions with (a-b)~price, (c-d)~rating as the ground truth segregated by explanation types. Overall for less than 10\% queries the bias score came out to be 0. The high width toward bias score 1 for the violin plot further suggests that lower priced and better rated products were available in the SERP, but were not added to cart by Alexa.
	}
	\label{Fig: Ground-Truth-1000}
	\vspace{-5 mm}
\end{figure}

\vspace{1 mm}
\noindent
\textbf{Bias score distribution with other ground truths: } Figure~\ref{Fig: Ground-Truth-1000} shows the CDF and violin plots for the bias score distributions with price and rating of the products being the ground truth. The higher mean bias scores mentioned in Table~\ref{Tab: BiasBreakUp} for these ground truths coupled with the different distributions suggest that for a very significant fraction of the queries the best priced and best rated products were not added to cart by Alexa. For less than 10\% of the queries the best rated and best priced products were added to cart. For the rest (more than) 90\% cases, a better priced and / or better rated product was available at a better position on the SERP; however the same was never added to cart by Alexa.
Note that, here any tie between products was broken by the relevance of position on the SERP (see Section~\ref{Sec: Deservingness} for details). 

\subsection{Results for temporal snapshots}~\label{Sec: Sup-temporal}
\begin{figure}[t]
	\centering
	\begin{subfigure}{0.48\columnwidth}
		\centering
		\includegraphics[width= \textwidth, height=3cm]{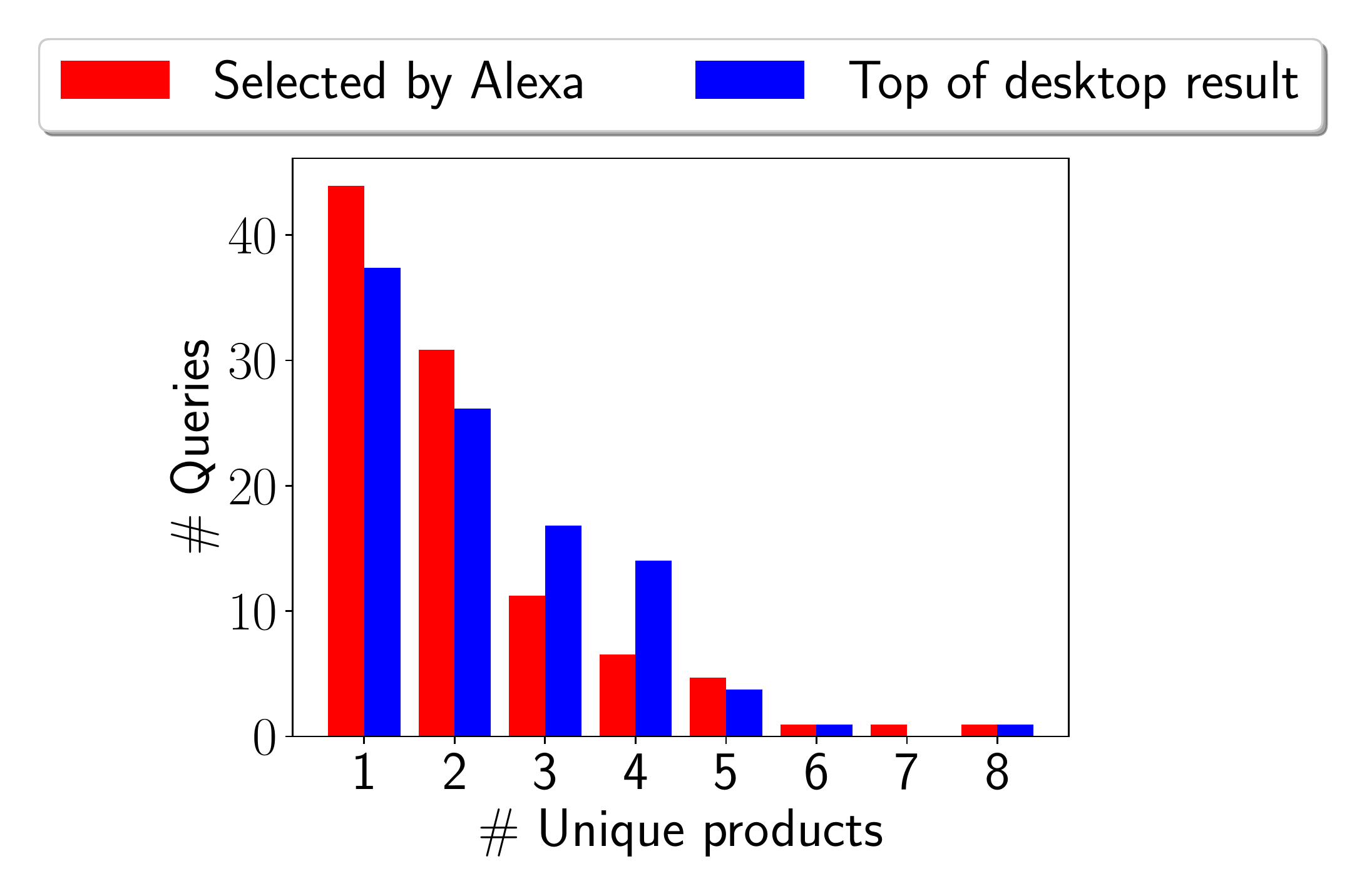}
		\caption{}
	\end{subfigure}
	\begin{subfigure}{0.48\columnwidth}
		\centering
		\includegraphics[width= \textwidth, height=3cm]{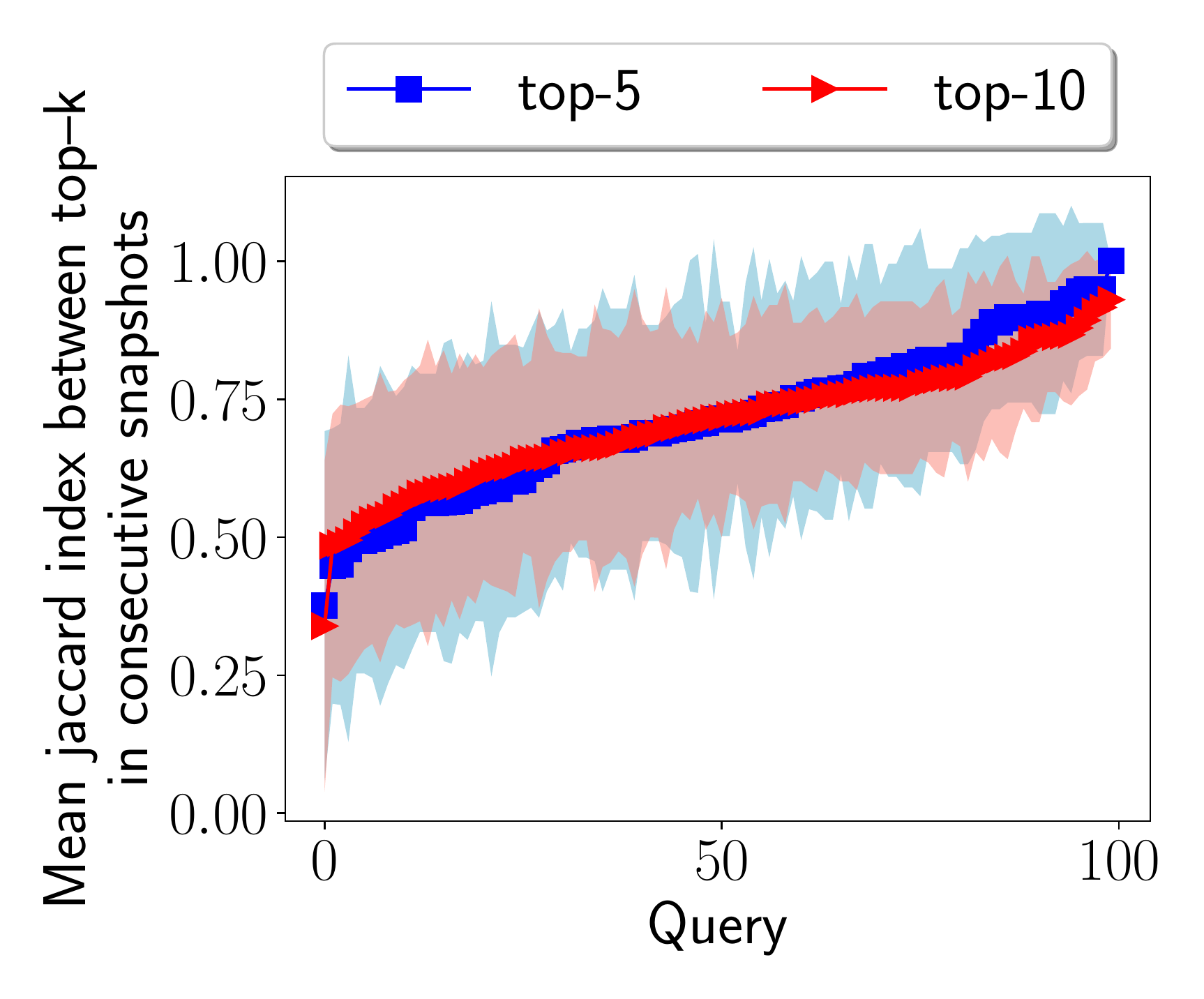}
		\caption{}
	\end{subfigure}
	\caption{ (a)~For more than 80\% of the queries the top of desktop search result and / or the product added to cart by Alexa were occupied by at most three unique ASINs throughout the 14 temporal snapshots. (b)~For more than 80\% of the queries more than 60\% products are retained across consecutive snapshots for top--5 and top--10 desktop search results. 
	}
	\label{Fig: Stability}
	\vspace{-5 mm}
\end{figure}

As mentioned in Table~\ref{Tab: BasicBreakUp}, more than 96\% of the instances (out of the 1400 times the queries were fired), the products were added to cart with Amazon's Choice and `a top result explanations'. In this section, we show results analogous to Figure~\ref{Fig: AC-Interpretation}, Table~\ref{Tab: BiasBreakUp} and Figures~\ref{Fig: Bias-Score-Distribution} and~\ref{Fig: Ground-Truth-1000} for the temporal snapshots with brief descriptions. In Figure~\ref{Fig: Stability}(a), we show the number of unique products added to cart or appeared as top of desktop search for different number of queries throughout the 14 temporal snapshots. In Figure~\ref{Fig: Stability}(b), we show the mean jaccard index (along with the standard deviation) between set of products in top--k desktop search results in two consecutive temporal snapshots for all the 100 queries. The mean jaccard index of the shwon distributions is 0.7 and 0.72 for top--10 and top--5 curves.

\begin{figure}[t]
	\centering
	\begin{subfigure}{0.48\columnwidth}
		\centering
		\includegraphics[width= \textwidth, height=3cm]{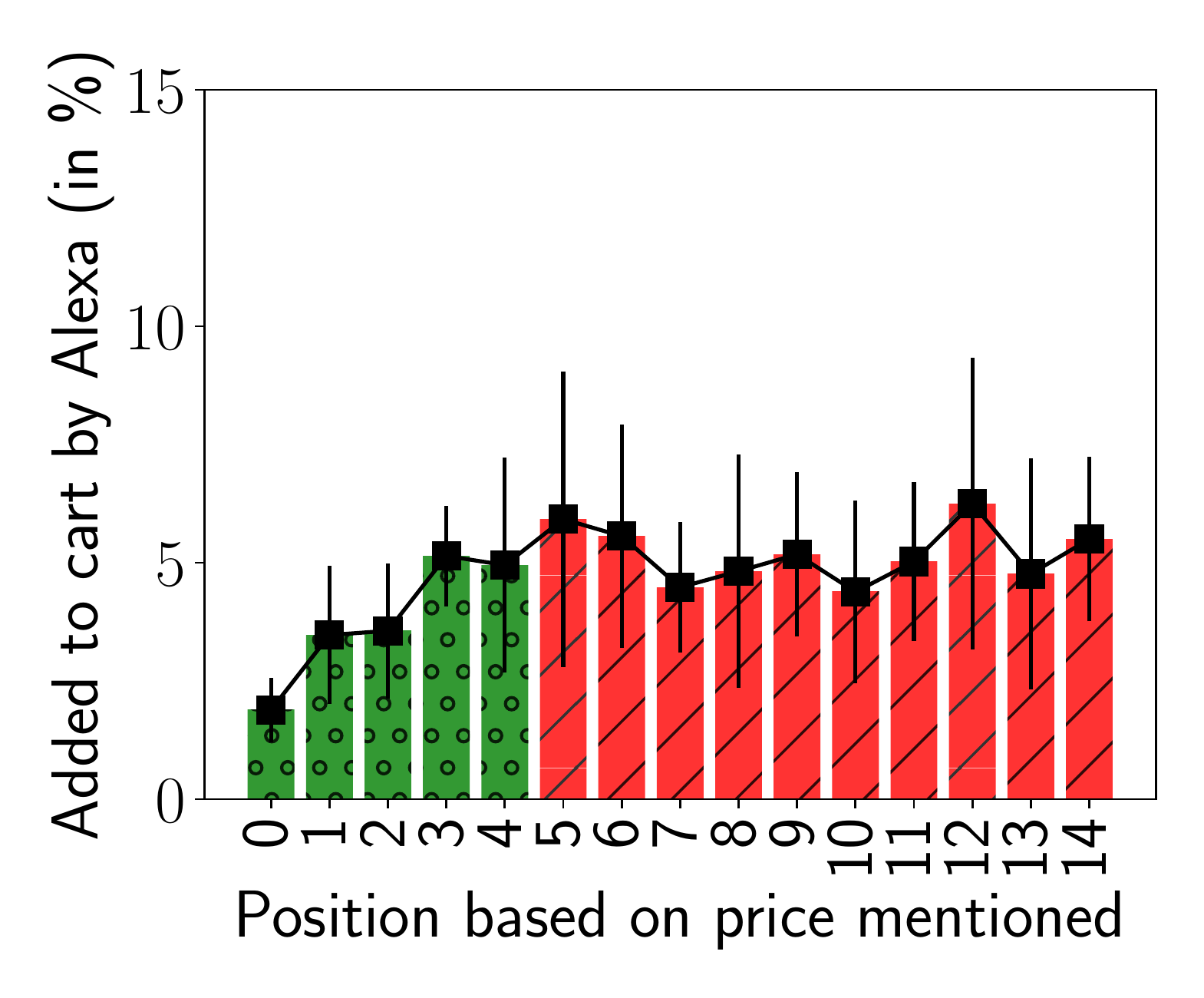}
		\caption{AC -- Price}
	\end{subfigure}
	\begin{subfigure}{0.48\columnwidth}
		\centering
		\includegraphics[width= \textwidth, height=3cm]{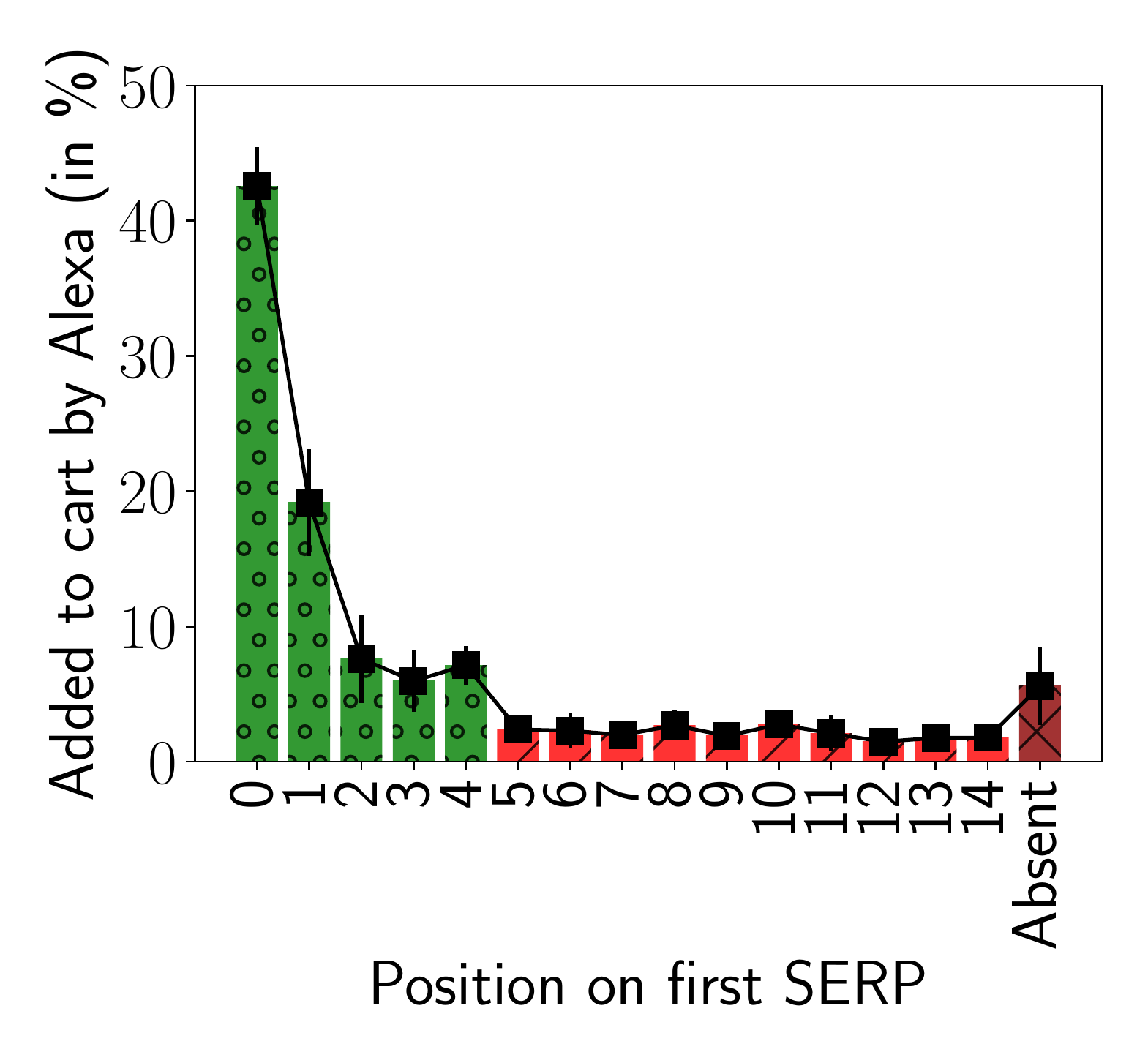}
		\caption{AC -- Position}
	\end{subfigure}
	
	\begin{subfigure}{0.48\columnwidth}
		\centering
		\includegraphics[width= \textwidth, height=3cm]{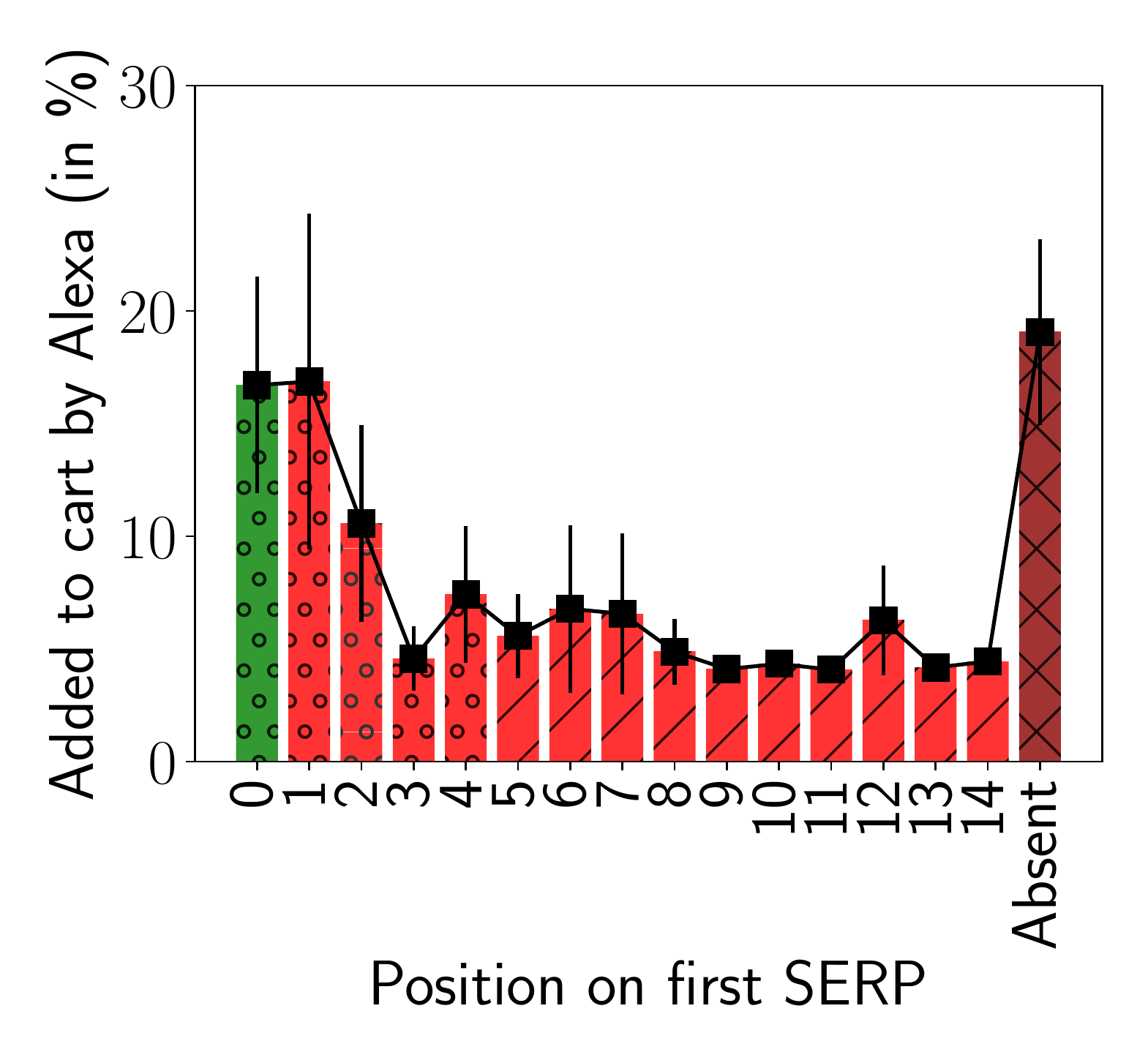}
		\caption{Top -- Position}
	\end{subfigure}
	\begin{subfigure}{0.48\columnwidth}
		\centering
		\includegraphics[width= \textwidth, height=3cm]{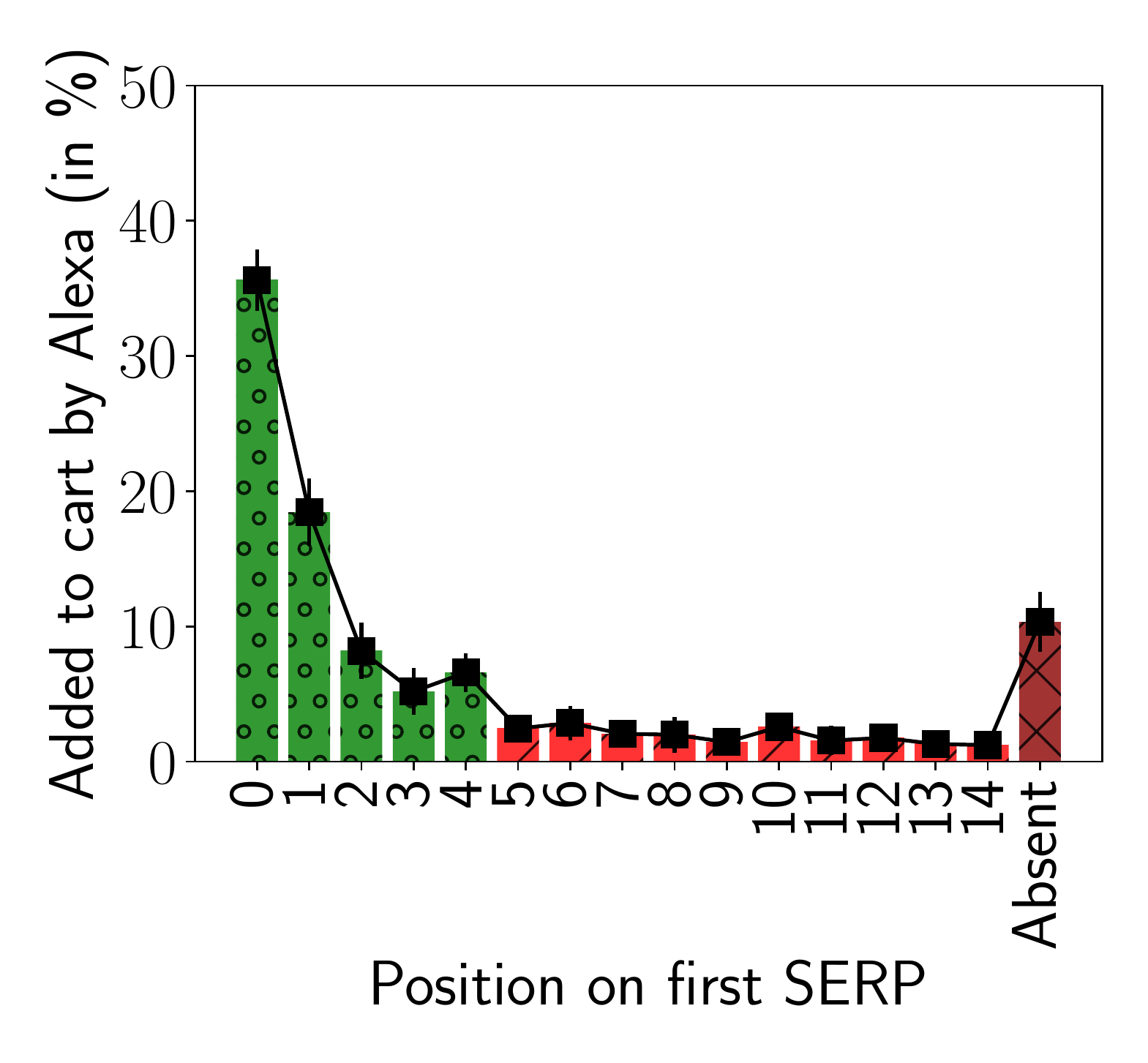}
		\caption{Overall -- Position}
	\end{subfigure}
	\caption{ The break ups (in \%) of rank of different products, as per their (a)~price mentioned and (b, c, and d)~position on first SERP, that were added to cart with different explanations. The distributions suggest that in significant number of cases, a better option was available than the product which was added to cart by Alexa. Green color in the figures indicate positions where interpretation and observations match.
	}
	\label{Fig: Temporal-Interpretation}
	\vspace{-5 mm}
\end{figure}

\vspace{1 mm}
\noindent
\textbf{Interpretations of explanations provided by Alexa:} Figure~\ref{Fig: Temporal-Interpretation} shows the rank wise break up of products added to cart by Alexa based on different explanations during the temporal snapshot data collection. The height of each bar indicates the mean of the contribution from each of the rank over all the 14 snapshots. Each bar is accompanied with the standard deviation error plot to show the deviation across different snapshots. Much like the 1000 query snapshot results (Figure~\ref{Fig: AC-Interpretation}), in these temporal snapshots we find that for a significant majority of the cases there existed at least one more relevant or less priced product which could have been added to cart by Alexa. While products added with Amazon's Choice explanations were highly rated as per the interpretations mentioned in Table~\ref{Tab: Summary-Takeaways}, they did not adhere to the interpretation of well priced products. For `a top result' based explanations, again less than 20\% products on average are actually from position 1 (which is the interpretation of top result according to 62\% respondents). Therefore, the interpretation of top-result does not align with the observation across temporal snapshots. \textit{These observations further underpins the significant gap in customers' interpretation of the explanation and the actual observation made from our 1000 query snapshot was not a pathological case}. Rather, across data collected over a period of two weeks similar trends were observed.

\begin{table}[tb]
	\noindent
	\small
	\centering
	\begin{tabular}{ |p {3.0 cm}| p {1.2 cm}|p {1.0 cm}|p {0.8 cm}||p {0.8 cm}|}
		\hline
		\bf Ground-truth &\bf Amazon's Choice & \bf A top result & \bf Others & \bf Overall\\
		\hline
		Based on position on SERP & 0.38 & 0.65 & 0.66 & 0.46\\
		\hline
		Based on price on SERP & 0.90 & 0.89 & 0.82 & 0.89\\ 
		\hline
		Based on rating on SERP & 0.83 & 0.80 & 0.86 & 0.82\\
		\hline
	\end{tabular}	
	\caption{ Mean bias scores due to product selection by Alexa based on the different mentioned ground truth rankings and the mentioned explanation type across different queries. 
	}
	\label{Tab: BiasBreakUp-Temporal}
	\vspace{-8 mm}
\end{table}

\begin{figure}[t]
	\centering
	\begin{subfigure}{0.48\columnwidth}
		\centering
		\includegraphics[width= \textwidth, height=3cm]{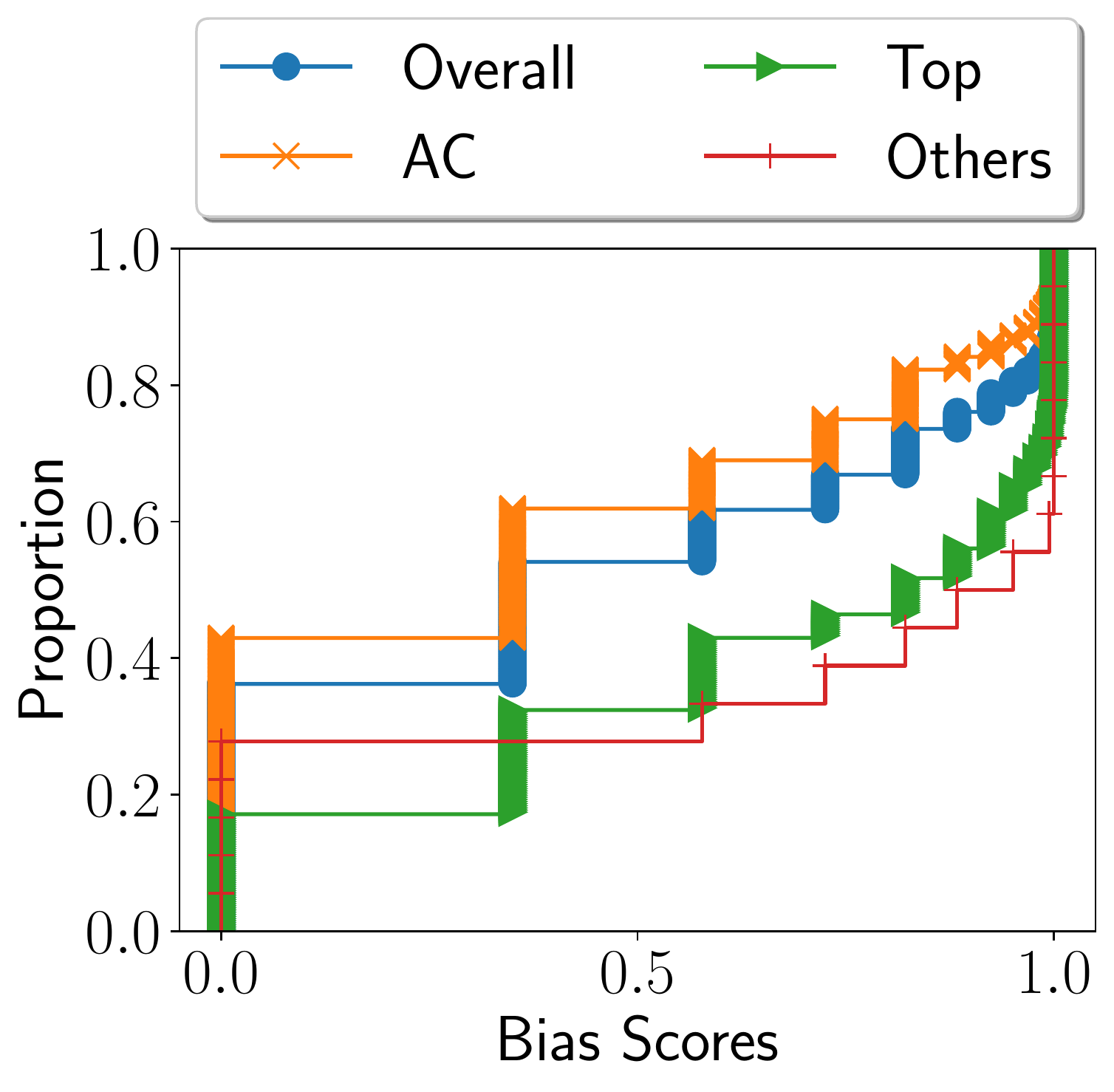}
		\caption{}
	\end{subfigure}
	\begin{subfigure}{0.48\columnwidth}
		\centering
		\includegraphics[width= \textwidth, height=3cm]{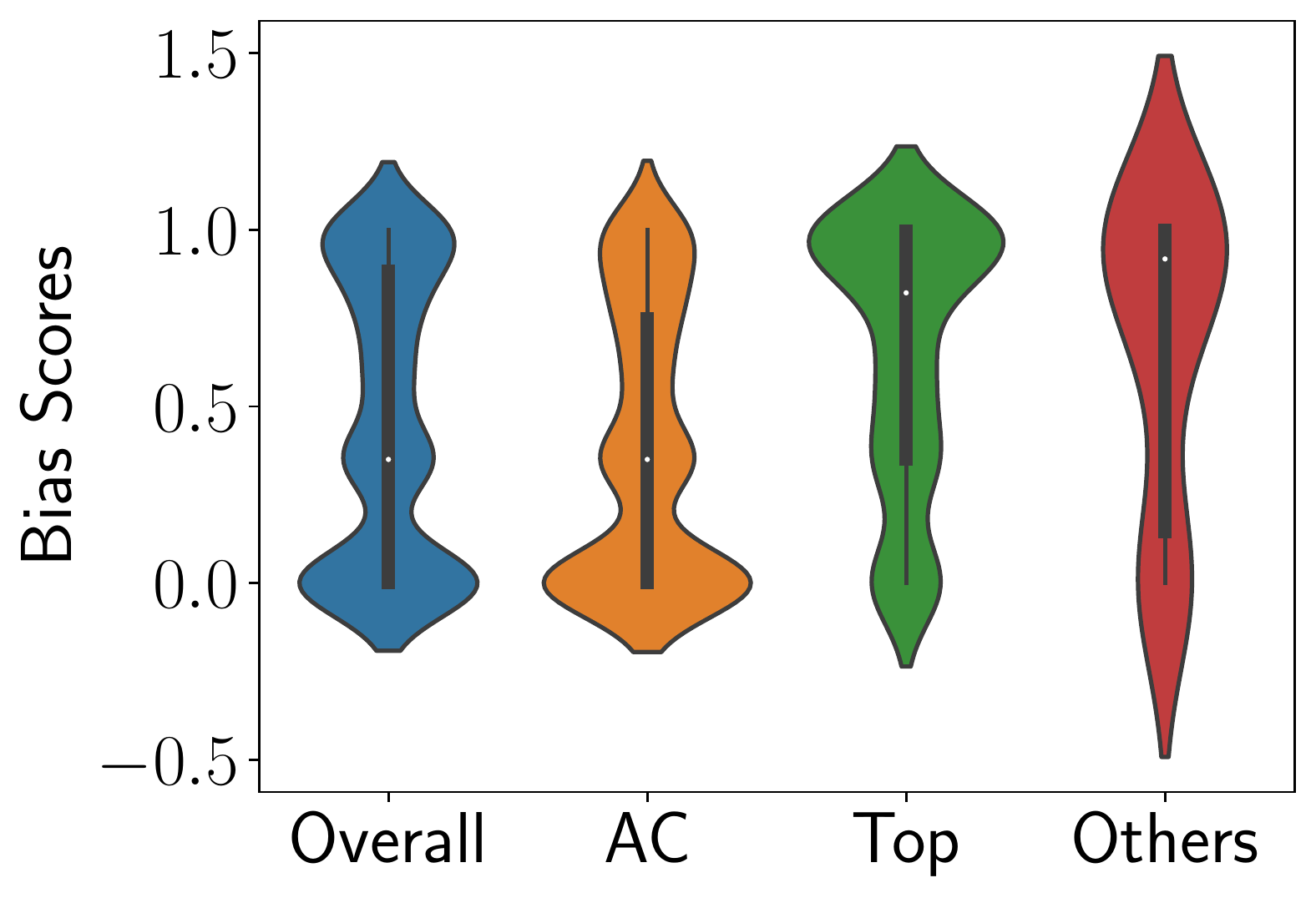}
		\caption{}
	\end{subfigure}
	
	\begin{subfigure}{0.48\columnwidth}
		\centering
		\includegraphics[width= \textwidth, height=3cm]{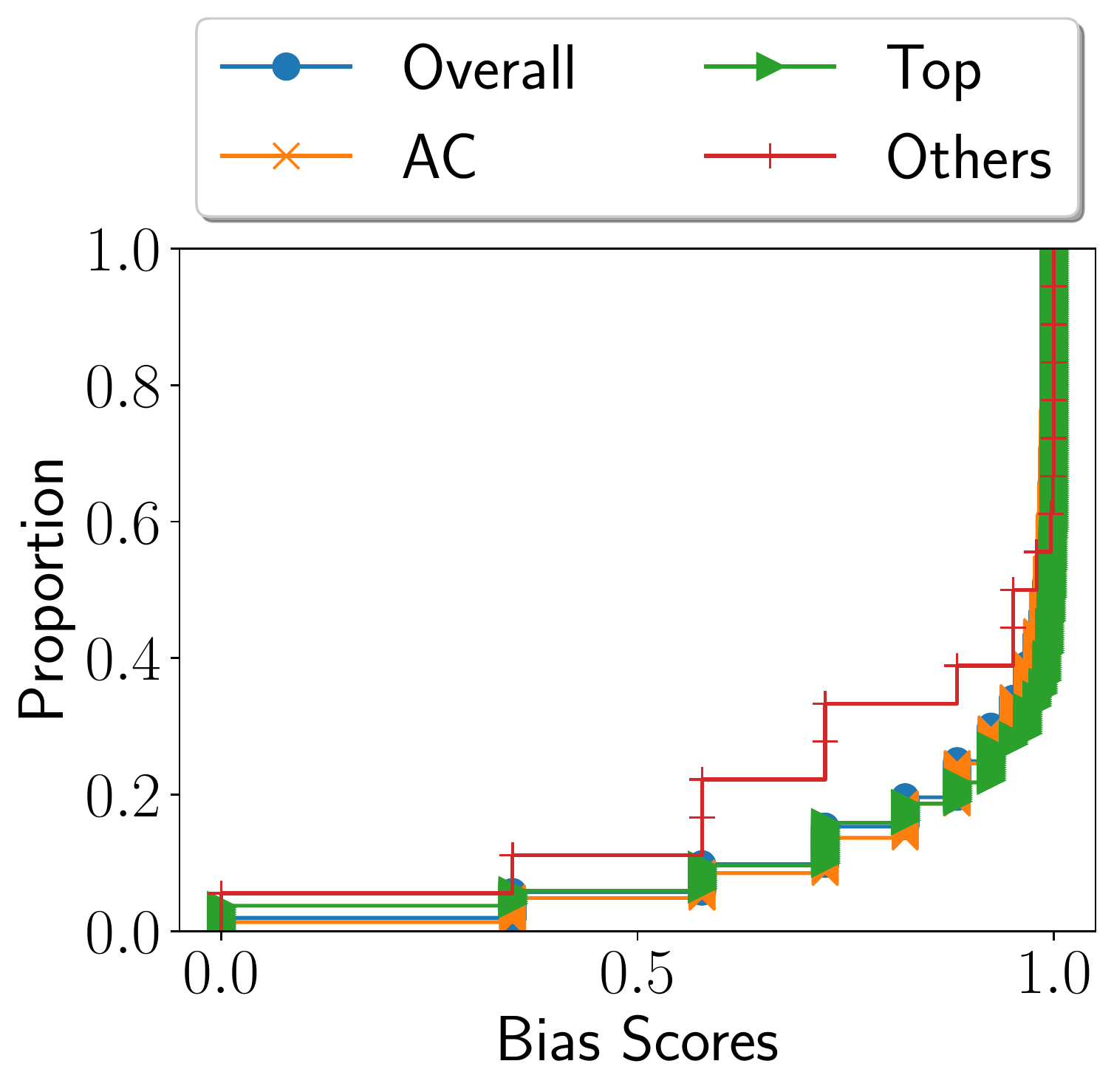}
		\caption{}
	\end{subfigure}
	\begin{subfigure}{0.48\columnwidth}
		\centering
		\includegraphics[width= \textwidth, height=3cm]{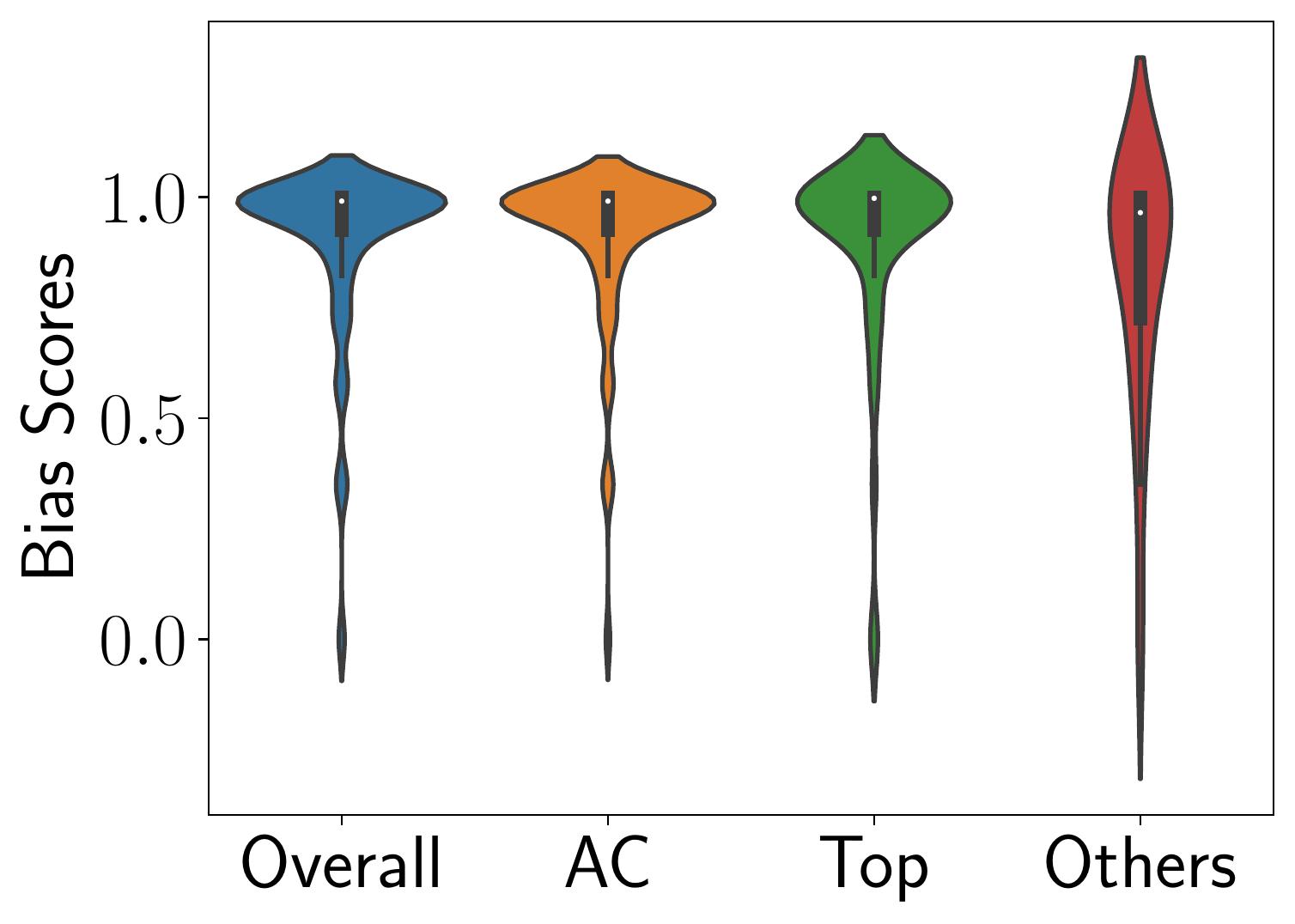}
		\caption{}
	\end{subfigure}

	\caption{ CDF and violin plot of the bias score distributions with (a--b)position, (c--d)~price, 
	as the ground truth segregated by explanation types. Overall for very less percentage of queries the bias score was evaluated to be 0. The high width toward bias score 1 for the violin plot further suggests that more relevant, and lower priced 
	products were available in the SERP but were not added to cart by Alexa.
	}
	\label{Fig: Temporal-Bias-Distributions}
	\vspace{-5 mm}
\end{figure}

\vspace{1 mm}
\noindent
\textbf{Fairness of the status quo action: }Table~\ref{Tab: BiasBreakUp-Temporal} shows the mean bias score across the 1400 instances segregated as per different explanations and on an overall basis. In general, the bias score with position on SERP ground truth is 0.05 less than what we observed in the 1000 query snapshot (Table~\ref{Tab: BiasBreakUp}). This may be attributed to these queries being very popular and therefore, we see a better consistency. Further, in 5\% more occasions products are added with Amazon's choice explanation which in general appear at better ranks, thus bringing down the overall bias scores. However, with respect to the other two ground truths the bias scores are worse. 

A closer look into the distributions of the bias scores, however paint a very similar qualitative picture as was seen in Figures~\ref{Fig: Bias-Score-Distribution} and~\ref{Fig: Ground-Truth-1000}. Figure~\ref{Fig: Temporal-Bias-Distributions} shows the CDF and violin plots of the distributions. Figures with rating distribution as ground truth has been omitted for brevity. Much like in the 1000 query snapshot (32\%), here also merely 35\% times the most relevant product was added to cart by Alexa (blue curve in Figure~\ref{Fig: Temporal-Bias-Distributions}(a)). Higher width toward bias score 1 in the violin plots further indicate that for considerable percentage of cases, the product added to cart was from a poor rank. The observation for products with `a top result' explanation is very similar to as noted in Figure~\ref{Fig: Bias-Score-Distribution}. These observations further corroborates with those made in Section~\ref{Sec: Deservingness}, i.e., \textit{for a significant percentage of instances the most relevant (or best priced or best rated) products were not added to cart by Alexa}. There exist at least one or more better option than the one added to cart as part of the status quo. This further highlights the apprehended unfairness concerns associated with the status quo action.

\end{document}